\documentclass[a4paper,final,11pt,twoside]{article}
\usepackage{amsfonts}

\newtheorem{theorem}{Theorem}[section]

\newtheorem{lemma}[theorem]{Lemma} 
\newtheorem{proposition}[theorem]{Proposition} 
\newtheorem{definition}[theorem]{Definition}

\newtheorem{remark}[theorem]{Remark}

\input{tcilatex}
\begin{document}

\author{G. Dell'Antonio$^{(1)(2)}$, G. Panati$^{(1)}\bigskip $ \\
{\small (1) SISSA-ISAS, International School for Advanced Studies, Trieste.}%
\\
{\small (2) Dipartimento di Matematica - Universit\`{a} di Roma ``La
Sapienza''.}}
\title{Zero-energy resonances and the flux -across-surfaces theorem\\
}
\date{\ }
\maketitle

\begin{abstract}
The flux-across-surfaces conjecture represents a corner stone in quantum
scattering theory because it is the key-assumption needed to prove the usual
relation between differential cross section and scattering amplitude. We
improve a recent result \cite{TDMB}\ by proving the conjecture also in
presence of zero-energy resonances or eigenvalues, both in point and
potential scattering.
\end{abstract}

\section{Introduction}

In the framework of quantum scattering theory (QST), experimental data and
theoretical predictions are usually compared by using the familiar formula
which relates the differential cross section and \ the scattering amplitude.

\noindent Focusing on the case of potential scattering, we notice that this
familiar formula follows rigorously $^{(}$\footnote{%
See for example [AJS], Ch. 7.}$^{)}$ from the following basic assumption:
the probability $P(\Sigma ,\Psi _{0})$ that a particle is detected by an
apparatus with active surface $\Sigma $ when the beam is prepared in the
state $\Psi _{0}$, is related to the asymptotic outgoing state $\Psi _{%
\mathrm{out}}$ by the formula 
\begin{equation}
P(\Sigma ,\Psi _{0})=\int_{C(\Sigma )}|\widehat{\Psi }_{\mathrm{out}}(%
\mathrm{k})|^{2}\,d^{3}\mathrm{k}  \label{Fundamental}
\end{equation}
where $C(\Sigma )=\{\lambda \mathrm{x}\in \Bbb{R}^{3}:\mathrm{x}\in \Sigma
,\lambda \geq 0\}$ is the cone generated by $\Sigma $ and the symbol $\quad 
\widehat{}\quad $ denotes Fourier transform. \ The problem of deducing (\ref
{Fundamental}) from more basic principles, is then a corner-stone in the
foundations of quantum scattering theory.\medskip

The first answer to this problem was \emph{Dollard's theorem} \cite{Do1}.
This theorem claims that, assuming existence and asymptotic completeness of
the wave operators $W_{\pm }=\lim_{t\rightarrow \pm \infty
}e^{iHt}e^{-iH_{0}t}$ (where $H_{0}=-\Delta $ with domain $H^{2}(\Bbb{R}%
^{3}) $ and $H=H_{0}+V$), one has

\[
\lim_{t\rightarrow \infty }\int_{C(\Sigma )}|\Psi _{t}(\mathrm{x}%
)|^{2}\,d^{3}\mathrm{x}=\int_{C(\Sigma )}|\widehat{\Psi }_{\mathrm{out}}(%
\mathrm{k})|^{2}\,d^{3}\mathrm{k} 
\]
where $\Psi _{t}=e^{-iHt}\Psi _{0}$ with $\Psi _{0}=W_{-}\Psi _{\mathrm{in}%
}=W_{+}^{*}\Psi _{\mathrm{out}}$. Dollard's theorem \ gives a possible
answer to the problem (\ref{Fundamental}) if one assumes that $P(\Sigma
,\Psi _{0})$ can be identified with the probability that the particle will
be (detected) in the cone $C(\Sigma )$ in the distant future.\mathstrut
\medskip

However, the experimental measurement is more closely related to the
probability that the particle is detected by the detector surface $\Sigma $
at \emph{any} time in the interval during which the detector is operating. \
This quantity can be computed by integrating the probability density current 
$\mathrm{j}^{\Psi _{t}}:=\func{Im}(\Psi _{t}^{*}\nabla \Psi _{t})$ on the
surface $\Sigma $ over the relevant time interval $(T_{1},T_{2})$. For the
foundations of QST, \ it is then relevant to prove that

\begin{equation}
\lim_{R\rightarrow \infty }\lim_{T_{2}\rightarrow \infty
}\int_{T_{1}}^{T_{2}}\,dt\int_{\Sigma _{R}}\mathrm{j}^{\Psi _{t}}\cdot 
\mathrm{n}\,d\sigma =\int_{C(\Sigma )}|\widehat{\Psi }_{\mathrm{out}}(%
\mathrm{k})|^{2}\,d^{3}\mathrm{k}  \tag*{(FAS)}  \label{FAS}
\end{equation}
where $\Sigma _{R}=C(\Sigma )\cap S_{R}$ and $S_{R}=\{\mathrm{x}\in \Bbb{R}%
^{3}:\left| \mathrm{x}\right| =R\}$.

This \emph{Flux-Across-Surfaces conjecture} was formulated by Combes, Newton
e Shtokhamer in 1975 \cite{CNS}. It was proved by Daumer, D\"{u}rr,
Goldstein, e Zangh\`{i} in \cite{DDGZ} for the case $V=0$ (see also \cite{DT}
for a simpler proof ). Then (FAS) has been proved for a wide class of \emph{%
short range } potentials by Amrein and Zuleta and \emph{long range}
potentials by Amrein and Pearson. \ In \cite{AP} and \cite{AZ} it is assumed
that $\widehat{\Psi }_{\mathrm{out}}$ has a compact support away from the
origin; such condition -- from a physical viewpoint -- can be seen as an 
\emph{energy cut-off }$.$ In \cite{TDMB} Teufel, D\"{u}rr and
M\"{u}nch-Berndl proved the FAS conjecture, for a different class of
potentials, without assuming this energy cut-off .

Despite the generality of these results, none of them relates to the case in
which there exists a zero-energy resonance or eigenvalue (for a rigorous
definition of zero-energy resonance see Definition \ref{Def ZER} and Sec.
4). Indeed, in \cite{TDMB} this eventuality it is explicitly ruled out in
the hypotheses. In \cite{AP} and \cite{AZ} it is assumed that $\widehat{\Psi 
}_{\mathrm{out}}$ has a compact support away from the origin and on this
(dense) class of states a zero-energy resonance is harmless. However, a
standard density argument cannot be applied, since the r.h.s of (FAS) is not
continuous in $\Psi _{0}$ (nor in $\Psi _{\mathrm{in}}$) in the $L^{2}$
topology.

In this paper we focus precisely on the case in which there exist a
zero-energy resonance for the pair $(H,H_{0})$ or \ zero is an eigenvalue
for the operator $H$ (we will refer to this situation as the \emph{resonant
case}). A zero-energy resonance affects the long-time behavior of the
wavefunction, and therefore the proof of FAS theorem.

\noindent Notice that the usual mapping properties of the (inverse) wave
operators -- which, roughly speaking, guarantee that the outgoing state $%
\Psi _{\mathrm{out}}$ inherits the same smoothness properties of the initial
state $\Psi _{0}$ -- fails to hold in the resonant case (see \cite{Ya}). \
In particular, we will show (see Proposition \ref{Prop reg}) that if there
exists a zero-energy resonance the asymptotic outgoing state generally has a
singularity in the origin of momentum space, in spite of the smoothness of
the initial state. \ This shows that in the resonant case it is not natural
at all to assume a smoothness condition on $\widehat{\Psi }_{\mathrm{out}}$,
as done in \cite{AP}, \cite{AZ} and \cite{TDMB} in the regular case.

\noindent It is interesting to notice that the scattering operator $%
S=W_{+}^{*}W_{-}$ maintains some nice mapping properties also in the
resonant case, as can be seen by using a stationary representation for $S$
(see, e.g. \cite{Agmon}, Th. 7.2 ) and our Proposition \ref{Prop reg}. \ \ 

\noindent However, we believe that the use of $\Psi _{\mathrm{in}}$ rather
than $\Psi _{0}$\ is somehow unphysical since the preparation procedure of
the system is performed at some past but finite time. In contrast the limit $%
T_{2}\rightarrow \infty $, which introduces the asymptotic state $\Psi _{%
\mathrm{out}}$, can be regarded as a reasonable approximation, convenient to
make connection with scattering theory. Since the FAS problem is deeply
related to its physical counterpart, we prefer to assume hypotheses only on
the initial state $\Psi _{0}$. This reflects, as we have pointed \ out, in
some additional mathematical troubles related to the singularity of $%
\widehat{\Psi }_{\mathrm{out}}$.\medskip \mathstrut

In Sec. 2 we prove (FAS) in the case of point interaction scattering.
Although this solvable model was already treated in \cite{PT} we present an
entirely different proof, using ideas which will be then applied to the
general case of potential scattering. \ In Sec. 3. we will treat the case of
potential scattering, by studying the behavior of the Lippman-Schwinger
eigenfunctions in the resonant case (see Proposition \ref{Prop reg}) and
using this result to prove the FAS theorem.

\noindent In Sec. 4 we summarize some previous results on zero-energy
resonances and we prove that, under suitable assumption on the potential,
some definitions of \ zero-energy resonance that can be found in the
literature (and that we use in Sec. 3) are indeed equivalent. Although this
is more or less common knowledge in scattering theory, we were not able to
find in the literature a result analogous to Proposition \ref{Prop
Equivalence}.\bigskip

\noindent \textbf{Acknowledgments.} It is a pleasure to thank Sandro Teta
for many helpful comments and remarks, and Detlef D\"{u}rr and \ Stefan
Teufel for valuable discussion during the preparation of this paper.\bigskip

\noindent \textbf{Convention. }We denote with $\mathcal{S}(\Bbb{R}^{d})$ the
Schwartz space of fast-decreasing smooth functions, with $\mathcal{S}%
^{\prime }(\Bbb{R}^{d})$ the space of tempered distributions and with $%
\left\langle \ldots ,\ldots \right\rangle $ the \emph{sesquilinear} pairing
between them. For any $u\in \mathcal{S}^{\prime }(\Bbb{R}^{d})$ we denote
its Fourier transform (resp. antitransform) as $\mathcal{F}u=\hat{u}$ \
(resp. $\mathcal{F}^{-1}u=\check{u}$). Derivatives and Fourier transforms
will be always intended in the sense of tempered distributions.\bigskip

\noindent \textbf{Convention. }Let be $\mathcal{E}$ \ a Banach space. We
denote as $\mathcal{B}(\mathcal{E})$ the algebra of the bounded operators on 
$\mathcal{E}$ and as $\mathcal{B}_{\infty }(\mathcal{E})$ the ideal of the
compact operators on $\mathcal{E}$. \ 

\section{A solvable example: point interaction scattering}

Point interaction is a widely used model to describe physical situations in
which a particle interacts with a potential whose range is negligible with
respect to the de Broglie wavelength of the particle. \ The point
interaction hamiltonian will be denoted as $H_{\gamma ,\mathrm{y}}$ where $%
\mathrm{y}\in \Bbb{R}^{3}$ denotes the point in which the interaction is
localized and $\gamma \in \Bbb{R}$ is a parameter related to the strength of
the interaction (in particular $\gamma =+\infty $ corresponds to the free
case).

From a mathematical point of view (see \cite{AGHH} and references therein) $%
H_{\gamma ,\mathrm{y}}$ can be rigorously defined by using the von
Neumann-Kre\u{\i}n extension theory. It is known that the continuous
spectrum is purely absolutely continuous and $\sigma _{\mathrm{ac}%
}(H_{\gamma ,\mathrm{y}})=[0,+\infty )$. The point spectrum is empty if $%
\gamma \leq 0$ and \ \ \ $\sigma _{\mathrm{p}}(H_{\gamma ,\mathrm{y}%
})=\left\{ -(4\pi \gamma )^{2}\right\} $\ \ for $\gamma >0.$ For $\gamma =0$
the hamiltonian exhibits a zero-energy resonance.

\noindent By using the generalized eigenfunctions 
\begin{equation}
\Phi _{\pm }(\mathrm{x},\mathrm{k})=e^{i\mathrm{k\cdot x}}+\frac{e^{i\mathrm{%
k\cdot y}}}{(4\pi \gamma \pm i|\mathrm{k}|)}\frac{e^{\mp i|\mathrm{k}||%
\mathrm{x-y}|}}{|\mathrm{x-y}|}
\end{equation}
one can define two unitary maps $\mathcal{F}_{\pm }:$ $\mathcal{H}%
_{ac}(H_{\gamma ,\mathrm{y}})\rightarrow L^{2}(\Bbb{R}^{3})$ by posing 
\[
(\mathcal{F}_{\pm }f)(\mathrm{k})=\stackunder{R\rightarrow +\infty }{\ 
\mathrm{s\ lim}}\int_{B_{R}}\Phi _{\pm }^{*}(\mathrm{x},\mathrm{k})f(\mathrm{%
x})\ (2\pi )^{-3/2}dx. 
\]
The operators $\mathcal{F}_{\pm }$ spectralize the operator $H_{\gamma ,%
\mathrm{y}}$ and are related to the wave operators $W_{\pm
}=\lim_{t\rightarrow \pm \infty }e^{H_{\gamma ,\mathrm{y}}t}e^{-iH_{0}t}$
(here $H_{0}=-\frac{1}{2}\Delta $ ) by the intertwining properties 
\begin{equation}
W_{\pm }^{-1}=\mathcal{F}^{-1}\mathcal{F}_{\pm }\qquad \text{and\qquad }%
W_{\pm }=\mathcal{F}_{\pm }^{-1}\mathcal{F}  \label{WOP relations}
\end{equation}
where $\mathcal{F}$ is the usual Fourier transform.

\noindent Since the operators $H_{\gamma ,\mathrm{y}}$ for different choice
of $\mathrm{y}\in \Bbb{R}^{3}$ are unitarly equivalent, in the following we
will consider only the case $\mathrm{y}=0$.\bigskip

As previously pointed out, when the hamiltonian exhibits a zero-energy
resonance (i.e. for $\gamma =0$), the asymptotic outgoing state $\Psi _{%
\mathrm{out}}$ is singular in momentum space, in spite of the smoothness of
the initial state $\Psi _{\mathrm{0}}$. This can be immediately seen by
noticing that 
\begin{eqnarray}
\widehat{\Psi }_{\mathrm{out}}(\mathrm{k}) &=&\left( \mathcal{F}_{+}\Psi
_{0}\right) (\mathrm{k})=\int_{\Bbb{R}^{3}}\Phi _{+}(\mathrm{x},\mathrm{k}%
)^{*}\Psi _{0}(\mathrm{x})\ (2\pi )^{-\frac{3}{2}}dx  \nonumber \\
&=&\widehat{\Psi }_{0}(\mathrm{k})+\frac{1}{4\pi \gamma -i|\mathrm{k}|}\int_{%
\Bbb{R}^{3}}\frac{e^{i|\mathrm{k}||\mathrm{x}|}}{|\mathrm{x}|}\Psi _{0}(%
\mathrm{x})\ (2\pi )^{-\frac{3}{2}}dx  \label{Psi out}
\end{eqnarray}
and recalling that the zero-energy resonance corresponds to $\gamma =0$.

However, in the solvable case of point interaction, it is possible to show
that $\widehat{\Psi }_{\mathrm{out}}$ has a smooth behavior outside the
origin and some decreasing behavior at infinity. More precisely, we can
prove the following lemma.

\begin{lemma}[Asymptotic decrease of the outgoing state]
Let be $\Psi _{\mathrm{0}}\in \mathcal{S}(\Bbb{R}^{3}).$ Let us define $\Psi
_{\mathrm{out}}:=W_{+}^{-1}\Psi _{\mathrm{0}}$ where $W_{\pm }$ are the wave
operators with respect to the pair $(H_{\gamma ,\mathrm{y}},H_{0})$ for $%
\gamma =0$. \label{Lemma Out state} Then $\widehat{\Psi }_{\mathrm{out}}\in
C^{\infty }(\Bbb{R}^{3}\backslash \{0\})$ and for every $m\in \Bbb{N}$ $\ $%
there exist positive constants $C_{m}$ and $K_{m}$ such that 
\begin{equation}
\left| \frac{\partial ^{m}}{\partial |\mathrm{k}|_{{}}^{m}}\widehat{\Psi }_{%
\mathrm{out}}(\mathrm{k})\right| \leq \frac{C_{m}}{|\mathrm{k}|^{3+m}}\qquad 
\text{for\qquad }\left| \mathrm{k}\right| \geq K_{m}.  \label{Out state 2}
\end{equation}
\end{lemma}

\noindent \textbf{Proof.} Since $\widehat{\Psi }_{0}\in \mathcal{S}(\Bbb{R}%
^{3})$, we can consider only the second term appearing in the second line of
equation (\ref{Psi out}) and we will denote it as $\zeta (k)$ where $k=|%
\mathrm{k}|$. By posing $\psi (x)=\int_{S_{1}}\Psi _{0}(x\omega )\ d\omega $%
, we get 
\begin{equation}
\zeta (k)=\frac{i}{k}\int_{0}^{+\infty }e^{ikx}x\psi (x)\ dx  \label{Csi}
\end{equation}
A dominated regularity argument immediately shows that $\zeta $ is $%
C^{\infty }$ for $k\in (0,+\infty )$. As for the behavior at infinity, we
notice that 
\begin{eqnarray*}
\zeta (k) &=&\frac{i}{k}\int_{0}^{+\infty }e^{ikx}\left( \psi (x)-\psi (0)\
e^{-x^{2}}-\psi ^{\prime }(0)\ xe^{-x^{2}}\right) x\ dx+ \\
&&+\frac{i}{k}\psi (0)\int_{0}^{+\infty }e^{ikx}\ e^{-x^{2}}x\ dx+\frac{i}{k}%
\psi ^{\prime }(0)\int_{0}^{+\infty }e^{ikx}\ e^{-x^{2}}x^{2}\ dx \\
&=&\zeta _{1}(k)+\zeta _{2}(k)+\zeta _{3}(k)
\end{eqnarray*}
where $\psi ^{\prime }$ is the right derivative of $\psi $ with respect to $%
x $. \ The second and third term \ can be computed exactly, getting that $%
\zeta _{2}(k)\asymp k^{-3}$ and $\zeta _{2}(k)\asymp k^{-4}$ as $%
k\rightarrow +\infty $. \ For the first term, we \ notice that 
\[
k^{3}\zeta _{1}(k)=-i\int_{0}^{+\infty }\left( \frac{d^{2}}{dx^{2}}%
e^{ikx}\right) \left( \psi (x)-\psi (0)\ e^{-x^{2}}-\psi ^{\prime }(0)\
xe^{-x^{2}}\right) \ dx. 
\]
By integrating by parts twice it follows that $\left| \zeta _{1}(k)\right|
\leq Ck^{-3}$. This proves (\ref{Out state 2}) for $m=0$. \noindent
\noindent By differentiating \ explicitly (\ref{Csi}) and using similar
arguments, one obtains (\ref{Out state 2}) for any $m\in \Bbb{N}$.%
\endproof%
\bigskip

Using this lemma, we can prove the FAS theorem for the point-interaction
scattering without making \textit{ad hoc} assumptions on the asymptotic
outgoing state. Although the statement is identical to [PT, Theorem 1] we
present a completely different proof, which can be generalized to the case
of potential scattering. With the previous notation, our result is the
following.

\begin{theorem}
Let us fix $\Psi _{0}\in \mathcal{S}(\Bbb{R}^{3})\cap \mathcal{H}%
_{ac}(H_{\gamma ,\mathbf{\mathrm{y}}})$. Then $\Psi _{t}:=e^{-iH_{\gamma ,%
\mathrm{y}}t}\Psi _{0}$ is continuously differentiable in $\Bbb{R}%
^{3}\backslash \{\mathbf{\mathrm{y}}\}$ and relation (FAS) holds true, for
every $T_{1}\in \Bbb{R}$. \label{Th pointFAS}
\end{theorem}

\noindent We focus on the proof in the resonant case $\gamma =0$ (see \cite
{PT} for $\gamma \neq 0$). To clarify the structure of the proof, we will
decompose it in some steps. To streamline the exposition, it is convenient
to introduce the following notation.

\begin{definition}
Fix $\nu \in \Bbb{R}_{+}$ $($We are interested in cases $\nu =\frac{1}{2}$
and $\nu =1)$. Consider an interval $[a,b]\subseteq \Bbb{R}$. We say that $F:%
\Bbb{R}^{3}\times \Bbb{R\rightarrow C}$ is of type $\mathcal{O}_{[a,b]}(%
\frac{|\mathrm{x}|}{t^{\nu }})$ if there exist $T_{0}>0$, $R_{0}>0$ such
that \label{Definition O} 
\begin{equation}
\stackunder{|\mathrm{x}|\geq R_{0},t\geq T_{0}}{\mathrm{Sup}}\left( \frac{|%
\mathrm{x}|}{t^{\nu }}\right) ^{\tau }\left| F(\mathbf{\mathrm{x}},t)\right|
\leq C_{\nu ,\tau }  \label{Omogen}
\end{equation}
for each $\tau \in [a,b]$. In this case we write $F=\mathcal{O}_{[a,b]}(%
\frac{|\mathrm{x}|}{t^{\nu }})$.\bigskip
\end{definition}

If $[a,b]=[0,n]$ (with $n\in \Bbb{N}$) we write, with a harmless abuse of
notation, $F=\mathcal{O}_{n}(\frac{|\mathrm{x}|}{t^{\nu }})$. The previous
definitions are trivially extended to the case of $\Bbb{C}^{d}$-valued
functions.\bigskip

\noindent \textbf{Preliminaries.} Since $\mathcal{F}_{+}$ spectralize the
hamiltonian $H_{\gamma ,\mathrm{y}}$ and $\mathcal{F}_{+}\Psi _{0}=\widehat{%
\Psi }_{\mathrm{out}}$ we get 
\begin{eqnarray*}
\Psi _{t}(\mathrm{x}) &=&\int_{\Bbb{R}^{3}}e^{-i\mathrm{k}^{2}t}\widehat{%
\Psi }_{\mathrm{out}}(\mathrm{k})\Phi _{+}(\mathrm{x},\mathrm{k})\,(2\pi
)^{-3/2}dk \\
&=&\int_{\Bbb{R}^{3}}e^{-i\mathrm{k}^{2}t}\widehat{\Psi }_{\mathrm{out}}(%
\mathrm{k})e^{i\mathrm{k}\mathbf{\cdot }\mathrm{x}}\,(2\pi )^{-3/2}dk+\int_{%
\Bbb{R}^{3}}e^{-i\mathrm{k}^{2}t}\widehat{\Psi }_{\mathrm{out}}(\mathrm{k})%
\frac{1}{i|\mathrm{k}|}\frac{e^{-i|\mathrm{k}||\mathrm{x}|}}{|\mathrm{x}|}%
(2\pi )^{-3/2}dk \\
&\equiv &\alpha (\mathrm{x},t)+\beta (\mathrm{x},t)\,.
\end{eqnarray*}
\noindent Then the probability density current is 
\begin{equation}
\mathrm{j}^{\Psi _{t}}=\func{Im}(\alpha ^{*}\nabla \alpha +\alpha ^{*}\nabla
\beta +\beta ^{*}\nabla \alpha +\beta ^{*}\nabla \alpha ).\,
\end{equation}
\noindent The first term $\mathrm{j}_{0}=\func{Im}(\alpha ^{*}\nabla \alpha
) $ corresponds to the free evolution of $\Psi _{\mathrm{out}}$, so using
the free flux-across-surfaces theorem \cite{DDGZ} one has 
\[
\lim_{R\rightarrow \infty }\int_{T}^{+\infty }dt\int_{\Sigma _{R}}\mathrm{j}%
_{0}(\mathrm{x},t)\mathbf{\cdot }\mathrm{n}\,d\sigma =\int_{C(\Sigma )}|%
\widehat{\Psi }_{\mathrm{out}}(\mathrm{k})|^{2}\,dk\,. 
\]
\noindent Therefore, to prove Theorem \ref{Th pointFAS} what remains to show
is that 
\begin{equation}
\lim_{R\rightarrow \infty }\int_{T}^{+\infty }dt\int_{\Sigma _{R}}|\mathrm{j}%
_{1}(\mathrm{x},t)\mathbf{\cdot }\mathrm{n}|\,d\sigma =0\,  \label{j1}
\end{equation}
where $\mathrm{j}_{1}:=\func{Im}(\alpha ^{*}\nabla \beta +\beta ^{*}\nabla
\alpha +\beta ^{*}\nabla \beta )$. In order to prove (\ref{j1}) we need
estimates on $\alpha ,\beta $ and their gradients.

\paragraph{Estimates on $\alpha $ and $\nabla \alpha $.}

First of all, we decompose $\alpha $ as $\alpha _{\mathrm{reg}}+\alpha _{%
\mathrm{sing}}$ by extracting the singular part of $\widehat{\Psi }_{\mathrm{%
out}}$, which can be read from (\ref{Psi out}). More precisely, we pose 
\begin{equation}
f_{1}(\mathrm{k}):=\widehat{\Psi }_{\mathrm{out}}(\mathrm{k})-\frac{r}{|%
\mathrm{k}|}e^{-|\mathrm{k}|^{2}}\qquad \text{where\qquad }r:=i\int_{\Bbb{R}%
^{3}}\frac{1}{|\mathrm{x}|}\Psi _{0}(\mathrm{x})\ dx  \label{f_1}
\end{equation}
and then we define 
\begin{eqnarray*}
\alpha _{\mathrm{reg}}(\mathrm{x},t) &=&\int_{\Bbb{R}^{3}}e^{i\mathrm{k}%
\mathbf{\cdot }\mathrm{x}}e^{-i\mathrm{k}^{2}t}\,f_{1}(\mathrm{k})(2\pi
)^{-3/2}dk \\
\alpha _{\mathrm{sing}}(\mathrm{x},t) &=&-\int_{\Bbb{R}^{3}}e^{i\mathrm{k}%
\mathbf{\cdot }\mathrm{x}}e^{-i\mathrm{k}^{2}t}\frac{r}{|\mathrm{k}|}e^{-|%
\mathrm{k}|^{2}}\,(2\pi )^{-3/2}dk.
\end{eqnarray*}
Moreover, 
\[
\nabla \alpha (\mathrm{x},t)=i\int_{\Bbb{R}^{3}}e^{i\mathrm{k}\mathbf{\cdot }%
\mathrm{x}}e^{-ik^{2}t}\mathrm{k}\widehat{\Psi }_{\mathrm{out}}(\mathrm{k}%
)\,(2\pi )^{-3/2}dk. 
\]

\noindent The properties of $\alpha _{\mathrm{reg}}$ and $\nabla \alpha $
are given in the following lemma, which will be useful also in the general
case of potential scattering.\bigskip

\begin{lemma}[Free evolution of a slow-decreasing state]
Let us suppose that $f\in C^{5}(\Bbb{R}^{3}\backslash \{0\})$ satisfies the
following assumptions:\label{FreeEvolution}

\begin{enumerate}
\item[a)]  \textbf{regularity in a neighborhood of the origin:} there exists
a suitable punctured neighborhood $U_{0}$ of the origin such that 
\begin{equation}
\partial ^{\mu }f\in L^{\infty }(U_{0})  \label{f hp1}
\end{equation}
for every multi-index $\ \mu \in \Bbb{N}^{3}$ with $1\leq |\mu |\leq 5;$

\item[b)]  \textbf{decrease at infinity:} for every multi-index $\ \mu \in 
\Bbb{N}^{3}$ with $1\leq |\mu |\leq 5$ there exists positive $C_{\mu }$ and $%
K_{\mu }$ such that 
\begin{equation}
\left| \partial ^{\mu }f(\mathrm{k})\right| \leq \frac{C_{\mu }}{|\mathrm{k}%
|^{3+|\mu |}}\qquad \text{for }\left| \mathrm{k}\right| \geq K_{\mu }.
\label{f hp2}
\end{equation}
\end{enumerate}

Then 
\begin{equation}
\alpha _{f}(\mathrm{x},t)\equiv \int_{\Bbb{R}^{3}}e^{i\mathrm{k}\cdot 
\mathrm{x}}e^{-i\mathrm{k}^{2}t}f(\mathrm{k})\ dk=\frac{1}{t^{3/2}}\mathcal{O%
}_{1}\left( \frac{|\mathrm{x}|}{t}\right) .  \label{FEclaim}
\end{equation}
\end{lemma}

\noindent \textbf{Proof of lemma.} Since $f\in L^{2}(\Bbb{R}^{3})$ as a
consequence of hypotheses (a)$_{\mu =0}$ and (b)$_{\mu =0}$, we obtain that 
\begin{equation}
\alpha _{f}(\mathrm{x},t)=(e^{-iH_{0}t}\check{f})(\mathrm{x})=\frac{e^{i%
\frac{\mathrm{x}^{2}}{2t}}}{(2it)^{3/2}}\int_{\Bbb{R}^{3}}\ e^{-i\frac{%
\mathbf{\mathrm{x}\cdot }\mathrm{y}}{t}}e^{i\frac{\mathrm{y}^{2}}{2t}}\check{%
f}(\mathrm{y})\ dy  \label{FEintegral}
\end{equation}
where $\check{f}$ denotes the Fourier antitransform of $f$. \ 

First of all, we show that $\check{f}$ \ satisfies the following properties: 
\begin{equation}
\left| \check{f}(\mathrm{y})\right| \leq \frac{C_{m}}{|\mathrm{y}|^{m}}%
\qquad \text{for }m=2,\ldots ,5\text{ and }\mathrm{y}\in \Bbb{R}^{3},
\label{FE1}
\end{equation}
\begin{equation}
\partial _{j}\ \check{f}\in L^{1}(\Bbb{R}^{3})\qquad \text{for every }%
j=1,\ldots ,3.  \label{FE2}
\end{equation}
The bound (\ref{FE1}) can be proved by observing that 
\begin{equation}
e^{i\mathrm{k}\cdot \mathrm{y}}=(-i)^{m}\frac{1}{|\mathrm{y}|^{m}}\left( 
\frac{\mathrm{y}}{|\mathrm{y}|}\cdot \nabla _{\mathrm{k}}\right) ^{m}e^{i%
\mathrm{k}\cdot \mathrm{y}}  \label{exp}
\end{equation}
so that (by integration by parts and a standard density argument) we get 
\begin{equation}
\check{f}(\mathrm{y})=i^{m}\frac{1}{|\mathrm{y}|^{m}}\int_{\Bbb{R}^{3}}e^{i%
\mathrm{k}\cdot \mathrm{y}}\left( \frac{\mathrm{y}}{|\mathrm{y}|}\cdot
\nabla _{\mathrm{k}}\right) ^{m}f(\mathrm{k})\ dk  \label{Fourier}
\end{equation}
Our assumptions imply that $\left( \frac{\mathrm{y}}{|\mathrm{y}|}\cdot
\nabla _{\mathrm{k}}\right) ^{m}f\in L^{1}(\Bbb{R}^{3})$ for $m=2,\ldots ,5$%
, so that (\ref{FE1}) follows from the Riemann-Lebesgue lemma.

A similar argument shows that $\left| \partial _{j}\ \check{f}(\mathrm{y}%
)\right| \leq \frac{C}{|\mathrm{y}|^{4}}$ for every $\mathrm{y}\in \Bbb{R}%
^{3}.$ \ As for the local behavior the fact that $k_{j}f\in L^{2}$ implies
that $\partial _{j}\check{f}\in L^{2}\subseteq L_{\mathrm{loc}}^{1}$ $.$
This completes the proof of (\ref{FE2}).\medskip

The claim (\ref{FEclaim}) means that for every $\tau \in [0,1]$ there exists 
$C_{\tau }>0$ such that 
\begin{equation}
\left( \frac{|\mathrm{x}|}{t}\right) ^{\tau }\left| \alpha _{f}(\mathbf{%
\mathrm{x}},t)\right| \leq \frac{C_{\tau }}{t^{3/2}}.  \label{FEclaim2}
\end{equation}
In the case $\tau =0$, by using the representation (\ref{FEintegral}) and
the bound (\ref{FE1}), we get 
\begin{eqnarray*}
\left| \alpha _{f}(\mathbf{\mathrm{x}},t)\right| &\leq &\frac{C}{t^{3/2}}%
\int_{\Bbb{R}^{3}}\left| \check{f}(\mathrm{y})\right| \ dy \\
&\leq &\frac{C}{t^{3/2}}\left\{ \int_{B_{1}(0)}\frac{C_{2}}{|\mathrm{y}|^{2}}%
\ dy+\int_{\Bbb{R}^{3}\backslash B_{1}(0)}\frac{C_{4}}{|\mathrm{y}|^{4}}\
dy\right\} \leq \frac{C^{\prime }}{t^{3/2}}
\end{eqnarray*}
In the case $\tau =1$, by using (\ref{FEintegral}) and (\ref{exp}) we obtain 
\[
\frac{|\mathrm{x}|}{t}\alpha _{f}(\mathbf{\mathrm{x}},t)=\frac{-ie^{i\frac{%
\mathrm{x}^{2}}{2t}}}{(4\pi it)^{3/2}}\int_{\Bbb{R}^{3}}\ e^{-i\frac{\mathbf{%
\mathrm{x}\cdot }\mathrm{y}}{t}}\left( \frac{\mathrm{x}}{|\mathrm{x}|}\cdot
\nabla \right) e^{i\frac{\mathrm{y}^{2}}{2t}}\check{f}(\mathrm{y})\ dy 
\]
and then 
\begin{equation}
\frac{|\mathrm{x}|}{t}\left| \alpha _{f}(\mathbf{\mathrm{x}},t)\right| \leq 
\frac{C}{t^{3/2}}\sum_{r=1}^{3}\int_{\Bbb{R}^{3}}\left| \frac{1}{t}y_{r}e^{i%
\frac{\mathrm{y}^{2}}{2t}}\check{f}(\mathrm{y})+e^{i\frac{\mathrm{y}^{2}}{2t}%
}\partial _{r}\check{f}(\mathrm{y})\right| \ dy  \label{FEfinal}
\end{equation}
As a consequence of (\ref{FE1}) and (\ref{FE2}) we have that $y_{r}\check{f}$
and $\partial _{r}\check{f}$ belong to $L^{1}(\Bbb{R}^{3})$, so (\ref
{FEfinal}) implies (\ref{FEclaim2}) in the case $\tau =1$. \ \noindent The
general case $\tau \in (0,1)$ follows by an interpolation argument. 
\endproof%
\bigskip

The previous lemma applies to $f_{1}(\mathrm{k})\,$given by (\ref{f_1}) and $%
f_{2}(\mathrm{k}):=\mathrm{k}\widehat{\Psi }_{\mathrm{out}}(\mathrm{k})$, so
that$\,$\ we get 
\begin{equation}
\alpha _{\mathrm{reg}}=\frac{1}{t^{3/2}}\mathcal{O}_{1}\left( \frac{|\mathrm{%
x}|}{t}\right) \qquad \text{and\qquad }\nabla \alpha =\frac{1}{t^{3/2}}%
\mathcal{O}_{1}\left( \frac{|\mathrm{x}|}{t}\right) .  \label{Alpha st1}
\end{equation}
Moreover, 
\begin{equation}
\alpha _{\mathrm{sing}}=\frac{1}{|\mathrm{x}|}\frac{1}{t^{1/2}}\mathcal{O}%
_{[-1,1]}\left( \frac{|\mathrm{x}|}{\sqrt{t}}\right)  \label{Alpha st2}
\end{equation}
Indeed, a straightforward computation gives 
\[
\alpha _{\mathrm{sing}}(\mathrm{x},t)=\frac{i\pi ^{3/2}}{|\mathrm{x}|\sqrt{%
1+it}}\varphi \left( \frac{i|\mathrm{x}|}{2\sqrt{1+it}}\right) 
\]
where $\varphi (z):=e^{z^{2}}\left( \mathrm{erfc}(z)-\mathrm{erfc}%
(-z)\right) $ for $z\in \Bbb{C}.$ To prove (\ref{Alpha st2}) it is then
sufficient to show that 
\begin{equation}
\stackunder{\arg (z)<\frac{3}{4}\pi }{\mathrm{Sup}}\left( z^{\tau }\varphi
(z)\right) \leq C_{\tau }  \label{Sup _mu}
\end{equation}
for every $\tau \in [-1,1]$. $^{(}$\footnote{%
Notice that, for $z=\frac{i|\mathrm{x}|}{2\sqrt{1+it}}$, one has $\arg (z)<%
\frac{3}{4}\pi $ for every value of $\mathrm{x}$ and $t$.}$^{)}$ For $\tau
\in [-1,0]$ the bound (\ref{Sup _mu}) is trivially true, since $\varphi $
has a first order zero in $z=0$ and is bounded at infinity in the specified
region. For $\tau \in (0,1]$, one notices that the asymptotic expansion of
the error function (in the specified region) assure that $z\varphi (z)$ is
bounded at infinity.

\paragraph{Estimates on $\beta $ and $\nabla \beta .$}

We turn now to the estimates on $\beta $ and $\nabla \beta $. As before, it
is convenient to extract the singular part of $\widehat{\Psi }_{\mathrm{out}%
} $. We pose 
\begin{equation}
f_{3}(\mathrm{k}):=\frac{1}{|\mathrm{k}|}\widehat{\Psi }_{\mathrm{out}}(%
\mathrm{k})-\frac{r}{|\mathrm{k}|^{2}}e^{-k^{2}}-\frac{c}{|\mathrm{k}|^{{}}}%
e^{-k^{2}}  \label{f_2}
\end{equation}
where $r\in \Bbb{C}$ has been defined in (\ref{f_1}), and $c\in \Bbb{C}$ is
the zeroth order term in the Laurent expansion of \ $\widehat{\Psi }_{%
\mathrm{out}}$; then we decompose $\beta $ as $\beta _{\mathrm{sing}%
,2}+\beta _{\mathrm{sing},1}+\beta _{\mathrm{reg}}$ where

\begin{eqnarray}
\beta _{\mathrm{sing},2}(\mathrm{x},t) &\equiv &\frac{-i}{|\mathrm{x}|}\int_{%
\Bbb{R}^{3}}e^{-i\mathrm{k}^{2}t}\frac{r}{|\mathrm{k}|^{2}}e^{-\mathrm{k}%
^{2}}e^{-i|\mathrm{k}||\mathrm{x}|}(2\pi )^{-3/2}dk  \label{Beta sing2} \\
\beta _{\mathrm{sing},1}(\mathrm{x},t) &\equiv &\frac{-i}{|\mathrm{x}|}\int_{%
\Bbb{R}^{3}}e^{-i\mathrm{k}^{2}t}\frac{c}{|\mathrm{k}|^{{}}}e^{-\mathrm{k}%
^{2}}e^{-i|\mathrm{k}||\mathrm{x}|}(2\pi )^{-3/2}dk  \label{Beta sing1} \\
\beta _{\mathrm{reg}}(\mathrm{x},t) &\equiv &\frac{-i}{|\mathrm{x}|}\int_{%
\Bbb{R}^{3}}e^{-i\mathrm{k}^{2}t}f_{2}(\mathrm{k})e^{-i|\mathrm{k}||\mathrm{x%
}|}(2\pi )^{-3/2}dk  \label{Beta reg}
\end{eqnarray}
The first two terms can be computed exactly by gaussian integration,
obtaining 
\begin{equation}
\beta _{\mathrm{sing},2}=\frac{1}{|\mathrm{x}|}\frac{1}{\sqrt{t}}\mathcal{O}%
_{1}\left( \frac{|\mathrm{x}|}{\sqrt{t}}\right) \qquad \text{and\qquad }%
\beta _{\mathrm{sing},1}=\frac{1}{|\mathrm{x}|}\frac{1}{t}\mathcal{O}%
_{2}\left( \frac{|\mathrm{x}|}{\sqrt{t}}\right) .  \label{Beta st1}
\end{equation}

As for the third term, we will show that 
\begin{equation}
\beta _{\mathrm{reg}}(\mathrm{x},t)\leq \frac{1}{|\mathrm{x}|}\frac{C}{|%
\mathrm{x}|+t}\qquad \text{for }|\mathrm{x}|>R_{0},t>T_{0}\text{ \ }
\label{Beta st2}
\end{equation}
for suitable $R_{0}$ and $T_{0}$. First of all, we pose $k=|\mathrm{k}|$ and 
$\tilde{f}(k):=\int_{\Bbb{S}^{2}}f_{3}(k\omega )\ d\omega $ getting 
\begin{equation}
\beta _{\mathrm{reg}}(\mathrm{x},t)=\frac{C}{|\mathrm{x}|}\int_{0}^{+\infty
}e^{-i\left( k^{2}t+k|\mathrm{x}|\right) }\tilde{f}(k)\ k^{2}dk.
\label{Beta integral}
\end{equation}
In order to apply a stationary phase method, we define 
\begin{equation}
\eta :=|\mathrm{x}|+t\text{\qquad and\qquad }\chi (k):=\frac{k^{2}t+k|%
\mathrm{x}|}{|\mathrm{x}|+t}  \label{Chi}
\end{equation}
observing moreover that 
\begin{equation}
\frac{1}{\chi ^{\prime }(k)}\leq \mathrm{Max}(1,k^{-1})\qquad \text{%
and\qquad }\frac{\chi ^{\prime \prime }(k)}{\chi ^{\prime }(k)^{2}}\leq 
\frac{1}{k}  \label{Chi st}
\end{equation}
where $\chi ^{\prime }$ indicate the derivative of $\chi $ with respect to $%
k $. From (\ref{Beta integral}) and definition (\ref{Chi}) it follows that

\[
\beta _{\mathrm{reg}}(\mathrm{x},t)=\frac{C^{\prime }}{|\mathrm{x}|\left( |%
\mathrm{x}|+t\right) }\int_{0}^{+\infty }\left( \frac{d}{dk}e^{-i\eta \chi
(k)}\right) \frac{1}{\chi ^{\prime }(k)}\tilde{f}(k)\ k^{2}dk 
\]
By recalling definition (\ref{f_2}) and using (\ref{Psi out}) and Lemma \ref
{Lemma Out state} it is easy to show that $\tilde{f}$ and $\frac{d\tilde{f}}{%
dk}$ belong to $C^{1}(0,+\infty )$, are bounded in a neighborhood of zero
and satisfy the bound 
\begin{equation}
\tilde{f}(k)\leq Ck^{-3}\qquad \text{and\qquad }\frac{d\tilde{f}}{dk}(k)\leq
Ck^{-4}  \label{f properties}
\end{equation}
for $k\rightarrow +\infty $. These facts imply that integration by part is
possible and that the boundary term is zero, so we get 
\begin{equation}
\left| \beta _{\mathrm{reg}}(\mathrm{x},t)\right| \leq \frac{C^{\prime }}{|%
\mathrm{x}|\left( |\mathrm{x}|+t\right) }\int_{0}^{+\infty }\left| \frac{d}{%
dk}\left( \frac{1}{\chi ^{\prime }(k)}\tilde{f}(k)\ k^{2}\right) \right| dk.
\label{Beta control}
\end{equation}

\noindent From (\ref{f properties}) and (\ref{Chi st}) it follows that the
integral appearing on the right-hand side of (\ref{Beta control}) is finite.
This proves (\ref{Beta st2}).

Finally, we give an estimate on $\nabla \beta $. By direct computation we
obtain 
\begin{eqnarray*}
\nabla \beta (\mathrm{x},t) &=&-\frac{1}{|\mathrm{x}|^{2}}\int_{\Bbb{R}%
^{3}}e^{-i\mathrm{k}^{2}t}e^{-i|\mathrm{k}||\mathrm{x}|}\widehat{\Psi }_{%
\mathrm{out}}(\mathrm{k})\frac{1}{i|\mathrm{k}|}(2\pi )^{-3/2}dk \\
&&+\frac{1}{|\mathrm{x}|}\frac{\mathrm{x}}{|\mathrm{x}|}\int_{\Bbb{R}%
^{3}}e^{-i\mathrm{k}^{2}t}e^{-i|\mathrm{k}||\mathrm{x}|}\widehat{\Psi }_{%
\mathrm{out}}(\mathrm{k})(2\pi )^{-3/2}dk \\
&\equiv &-\frac{1}{|\mathrm{x}|}\beta (\mathrm{x},t)+\left( \nabla \beta
\right) _{\mathrm{r}}(\mathrm{x},t)
\end{eqnarray*}
The term $\left( \nabla \beta \right) _{\mathrm{r}}$ can be treated exactly
as $\beta $; however, since the second order pole does not appear, we get
the bound 
\begin{equation}
\left( \nabla \beta \right) _{\mathrm{r}}\leq \frac{1}{|\mathrm{x}|}\frac{1}{%
t}\mathcal{O}_{2}\left( \frac{|\mathrm{x}|}{\sqrt{t}}\right) +\frac{1}{|%
\mathrm{x}|}\frac{C}{|\mathrm{x}|+t}.  \label{Beta st3}
\end{equation}
\bigskip

\noindent \textbf{Proof of Theorem \ref{Th pointFAS}.} We remarked that in
order to prove Theorem \ref{Th pointFAS} it is sufficient to prove (\ref{j1}%
). To achieve the proof, we notice that the singular term 
\[
-\frac{1}{|\mathrm{x}|}\beta _{\mathrm{sing},2}^{*}(\mathrm{x},t)\beta _{%
\mathrm{sing},2}(\mathrm{x},t) 
\]
appearing in $\beta \nabla \beta $ is \emph{real}, so\emph{\ }it does not
contribute to $\func{Im}(\beta ^{*}\nabla \beta ).$ As for all the remaining
terms, they can be shown to vanish as $R\rightarrow +\infty $ by using
estimates (\ref{Alpha st1}), (\ref{Alpha st2}), (\ref{Beta st1}), (\ref{Beta
st2}) and (\ref{Beta st3}). As an example we show how to prove that 
\begin{equation}
\lim_{R\rightarrow \infty }\int_{T}^{+\infty }dt\int_{\Sigma _{R}}|\func{Im}%
(\beta ^{*}\nabla \alpha )\mathbf{\cdot }\mathrm{n}|\,d\sigma =0\,.
\label{Vanishing}
\end{equation}
We observe that 
\begin{eqnarray*}
\int_{\Sigma _{R}}|\func{Im}(\beta ^{*}\nabla \alpha )\mathbf{\cdot }\mathrm{%
n}|\,d\sigma &\leq &4\pi R^{2}\left| \beta _{\mathrm{sing},2}+\beta _{%
\mathrm{sing},1}+\beta _{\mathrm{reg}}\right| \left| \nabla \alpha \right| \\
&\leq &4\pi R\left\{ \frac{1}{\sqrt{t}}\mathcal{O}_{1}\left( \frac{R}{\sqrt{t%
}}\right) +\frac{1}{t}\mathcal{O}_{2}\left( \frac{R}{\sqrt{t}}\right) +\frac{%
C}{R+t}\right\} \left\{ \frac{1}{t^{3/2}}\mathcal{O}_{1}\left( \frac{R}{t}%
\right) \right\}
\end{eqnarray*}
where we used (\ref{Alpha st1}), (\ref{Beta st1}) and (\ref{Beta st3}) by
identifying $R=|\mathrm{x}|$. Now one makes use of property (\ref{Omogen})\
with suitable choices of $\tau $ and $\nu $ in order to control the previous
expression. For example, the first term is 
\[
\frac{R}{\sqrt{t}}\mathcal{O}_{1}\left( \frac{R}{\sqrt{t}}\right) \frac{1}{%
R^{\varepsilon }t^{3/2-\varepsilon }}\frac{R^{\varepsilon }}{t^{\varepsilon }%
}\mathcal{O}_{1}\left( \frac{R}{t}\right) \leq \frac{C}{R^{\varepsilon
}t^{3/2-\varepsilon }} 
\]
for every $\varepsilon \in (0,\frac{1}{2})$. By similar computations we get 
\[
\int_{\Sigma _{R}}|\func{Im}(\beta ^{*}\nabla \alpha )\mathbf{\cdot }\mathrm{%
n}|\,d\sigma \leq \frac{C}{R^{\varepsilon }t^{3/2-\varepsilon }}+\frac{%
C^{\prime }}{Rt^{3/2}} 
\]
This bound is sufficient to prove the vanishing of\ the left-hand side of (%
\ref{Vanishing}) by applying the dominated convergence theorem. This
completes the proof of Theorem \ref{Th pointFAS} 
\endproof%

\newpage

\section{Zero energy resonances and the FAS theorem in potential scattering}

In this section and the following one we will consider the scattering theory
for the pair $(H,H_{0})$ where $H_{0}=-\Delta $ on the domain $\mathcal{D}%
(H_{0})=H^{2}(\Bbb{R}^{3})$ and $H=H_{0}+V$. Later on, we will focus on
potentials satisfying the following assumptions.

\begin{definition}
\label{Def_Ikebe}We say that a measurable function $V:\Bbb{R}^{3}\rightarrow 
\Bbb{R}$ belongs to the \emph{Ikebe class} $\mathrm{(I)}_{n}$ $($with $n\in 
\Bbb{N})$, if:

\begin{enumerate}
\item  $V$ is locally H\"{o}lder continuous except that in a finite number
of points

\item  $V\in L^{2}(\Bbb{R}^{3})$

\item  there exist $R_{0}>0$ and $\varepsilon >0$ such that $\left| V(%
\mathrm{x})\right| \leq \frac{C_{0}}{|\mathrm{x}|^{n+\varepsilon }}$\ for $|%
\mathrm{x}|\geq R_{0}$.
\end{enumerate}

\noindent Moreover, we define $\mathrm{(I)}_{\infty }=\bigcap_{n\in \Bbb{N}}%
\mathrm{(I)}_{n}$.
\end{definition}

\noindent The terminology follows from the fact that, for $n=2,$ these are
the hypotheses under which Ikebe's eigenfunction expansion theorem \cite
{Ikebe} has been proved. Under these assumptions, the operator $H$ is
self-adjoint on $\mathcal{D}(H_{0})$. Moreover $H$ has neither positive
eigenvalues nor singular continuous spectrum and $\sigma _{\mathrm{ac}%
}(H)=[0,+\infty )$. Finally, the wave operators $W_{\pm }=\lim_{t\rightarrow
\pm \infty }e^{iHt}e^{-iH_{0}t}$ exist and are asymptotically
complete.\medskip

In what follows the Laplace operator $\Delta $ will be intended to act on
the space of tempered distributions $\mathcal{S}^{\prime }(\Bbb{R}^{d})$.
The operator $-\Delta +\kappa ^{2}$ ($\kappa \in \Bbb{C}_{+}$), seen as an
operator in $\mathcal{S}^{\prime }(\Bbb{R}^{d})$, has a \emph{right} inverse 
$G_{\kappa }$ given explicitly by the convolution (in the sense of tempered
distributions) with $\tilde{G}_{\kappa }(\mathrm{x})=\frac{e^{i\kappa |%
\mathrm{x}|}}{|\mathrm{x}|}$.

\subsection{Lippman-Schwinger eigenfunctions and zero energy resonances}

The main tool of our analysis will be the fact that the operator $H$ can be
``diagonalized'' by means of the so-called Lippman-Schwinger eigenfunctions,
or generalized eigenfunction. \ The classical results concerning this
generalized eigenfunctions, proved in \cite{Ikebe} and \cite{Povzner}, are
summarized in \cite{TDMB}. Here we point out few basic facts.\medskip

\noindent If \ $V\in \mathrm{(I)}_{2}$, then for every $\mathrm{k}\in \Bbb{R}%
^{3}\setminus \{0\}$ the generalized eigenfunction $\Phi _{\pm }(\cdot ,%
\mathrm{k}):\Bbb{R}^{3}\rightarrow \Bbb{C}$ is defined as the unique \emph{%
continuous} solution of the Lippman-Schwinger (LS) equation 
\begin{equation}
\Phi _{\pm }(\mathrm{x},\mathrm{k})=e^{i\mathrm{k\cdot x}}-\frac{1}{4\pi }%
\int_{\Bbb{R}^{3}}\frac{e^{\mp i|\mathrm{k}||\mathrm{x}-\mathrm{y}|}}{|%
\mathrm{x}-\mathrm{y}|}V(\mathrm{y})\Phi _{\pm }(\mathrm{y},\mathrm{k})\ \
dy\qquad  \tag{LS}  \label{LS original}
\end{equation}
which satisfies the asymptotic condition $\lim_{|\mathrm{x}|\rightarrow
\infty }\left( \Phi _{\pm }(\mathrm{x},\mathrm{k})-e^{i\mathrm{k}\cdot 
\mathrm{x}}\right) =0$. In more abstract terms$^{(}$\footnote{%
Here and in the following we will consider only the upper sign in $\Phi
_{\pm }$, omitting pedices.}$^{)}$, the function $\eta _{\mathrm{k}}(\mathrm{%
x}):=\Phi _{+}(\mathrm{x},\mathrm{k})-e^{i\mathrm{k}\cdot \mathrm{x}}$ is
the unique solution of the equation

\begin{equation}
(1+G_{|\mathrm{k}|}V)\eta _{\mathrm{k}}=g_{\mathrm{k}}  \label{LS_cont}
\end{equation}
in $C_{\infty }(\Bbb{R}^{3})$, the space of continuous functions vanishing
at infinity, where $g_{\mathrm{k}}$ is given by 
\[
g_{\mathrm{k}}(\mathrm{x})=-\frac{1}{4\pi }\int_{\Bbb{R}^{3}}\frac{e^{-i|%
\mathrm{k}||\mathrm{x}-\mathrm{y}|}}{|\mathrm{x}-\mathrm{y}|}V(\mathrm{y}%
)e^{i\mathrm{k}\cdot \mathrm{y}}\ dy. 
\]

However, for our purposes, the choice of $C_{\infty }(\Bbb{R}^{3})$ is not
the most suitable one. As pointed out by Agmon and Kuroda (see e.g. \cite
{Agmon}) a convenient alternative topology for setting the LS equation is
given by the weighted Sobolev spaces, defined by

\[
H^{m,s}(\Bbb{R}^{d}):=\left\{ u\in \mathcal{S}^{\prime }(\Bbb{R}%
^{d}):\left\| \left( 1+|x|^{2}\right) ^{\frac{s}{2}}\left( 1-\Delta \right)
^{\frac{m}{2}}u\right\| _{L^{2}}<+\infty \right\} . 
\]
By varying the indexes $m,s\in \Bbb{R}$ one gets a net of spaces (in the
picture $s,s^{\prime }\geq 0$)

$
\begin{array}{ccccccccc}
&  &  &  &  &  &  &  &  \\ 
\subseteq & H^{2,s} & \subseteq & H^{1,s} & \subseteq & L_{s}^{2} & \subseteq
& H^{-1,s} & \subseteq \\ 
& \cap &  & \cap &  & \cap &  & \cap &  \\ 
\subseteq & H^{2} & \subseteq & H^{1} & \subseteq & L^{2} & \subseteq & 
H^{-1} & \subseteq \\ 
& \cap &  & \cap &  & \cap &  & \cap &  \\ 
\subseteq & H^{2,-s^{\prime }} & \subseteq & H^{1,-s^{\prime }} & \subseteq
& L_{-s^{\prime }}^{2} & \subseteq & H^{-1,-s^{\prime }} & \subseteq \\ 
&  &  &  &  &  &  &  & 
\end{array}
$

\noindent where one recognizes the usual weighted $L^{2}$-spaces $%
L_{s}^{2}\equiv H^{0,s}$ and the usual Sobolev spaces $H^{m}$ $\equiv
H^{m,0} $. Moreover we define the spaces 
\[
H^{m,s-0}:=\bigcap_{r<s}H^{m,r}\qquad \text{and\qquad }H^{m,s+0}:=%
\bigcup_{r>s}H^{m,r} 
\]
regarded as linear spaces. \medskip

The main advantage of this approach is that, if \ $G_{\kappa }=(-\Delta
+\kappa ^{2})^{-1}$ ($\kappa \in \Bbb{C}_{+}$) is regarded as an operator
between weighted Sobolev spaces, then the map $\kappa \mapsto G_{\kappa }$
can be continuously extended to the closed upper half-plane $\overline{\Bbb{C%
}}_{+}$ $.$

\noindent Notice that if \ $G_{\kappa }$ ($\kappa \in \Bbb{C}_{+}$) is
regarded as an element of \ $\mathcal{B}(L^{2},H^{2})$ it coincides with the
free resolvent $R_{0}(\kappa )=(H_{0}-\kappa ^{2})^{-1}\in \mathcal{B}%
(L^{2}) $ but -- in such a case -- the map $\kappa \mapsto R_{0}(\kappa )$ 
\emph{cannot} be continuously extended to $\overline{\Bbb{C}}_{+}$, since $%
\kappa \in \Bbb{R}$ implies $\kappa ^{2}\in \sigma (H_{0})$.

\noindent The following lemma -- which has been proved in \cite{JK} -- makes
precise the previous statements.

\begin{lemma}[Extension of the free resolvent]
Assume $s,s^{\prime }>\frac{1}{2}$ $,$ $s+s^{\prime }>2$ and $m\in \Bbb{Z}$.
Then $\kappa \mapsto G_{\kappa }$, considered as a $\mathcal{B}%
(H^{m-2,s},H^{m,-s^{\prime }})$-valued function, can be \emph{continuously}
extended to the region $\kappa \in \overline{\Bbb{C}}_{+}$.\label%
{Lemma Ext res}
\end{lemma}

The previous lemma allow us to give a meaning to $1+G_{\kappa }V$, as an
operator between weighted Sobolev spaces, also for $\kappa $ on the real
axis. To this end, we consider potentials satisfying the following
condition: 
\begin{equation}
V\text{ \ is a compact operator from }H^{m,0}\text{ to }H^{m-2,\beta }\text{
\ for some }\beta >2.  \tag{V$.m.\beta $}  \label{V.beta}
\end{equation}
This condition implies that $V$ is a compact operator from $\ H^{m,s}$ to $%
H^{m-2,s+\beta }$ for every $s\in \Bbb{R}$ since $V$ commutes with
multiplication by $\left( 1+|x|^{2}\right) ^{\frac{s}{2}}$. If \ $V\in 
\mathrm{(I)}_{n}$ then condition (V.$m.\beta $) holds true for $m=1,2$ for
every $\beta <n+\varepsilon $ (with $\varepsilon $ from Def. \ref{Def_Ikebe}%
.iii) (for $m=2$ this is a consequence of the fact that $\left\langle \cdot
\right\rangle ^{-\beta }V$ is an $H_{0}$-compact operator in $L^{2}$; for $%
m=1$ see \cite{JK}).

Assume condition (V.$m.\beta $) for $m=1,2$ \ and $\beta >2.$ Then, for
every $s>\frac{1}{2}$ and $m=1,2$ one has a compact-operator-valued analytic
map \ $\Bbb{C}_{+}\rightarrow \mathcal{B}_{\infty }(H^{m,-s}),\kappa \mapsto
G_{\kappa }V$ that, in virtue of Lemma \ref{Lemma Ext res}, can be extended
by continuity to the (positive) real axis. This allows us to formulate the
LS equation in $H^{m,-s}.$ The comparison with the previous approach is
given in the following proposition.

\begin{proposition}
Let be $s>\frac{1}{2}$ . Assume that $V\in \mathrm{(I)}_{2}$. Then, for
every $\mathrm{k}\in \Bbb{R}^{3}\setminus \{0\}$, there exists a unique
solution\ $\tilde{\eta}_{\mathrm{k}}$\ of \ the equation 
\begin{equation}
(1+G_{|\mathrm{k}|}V)\tilde{\eta}_{\mathrm{k}}=g_{\mathrm{k}}
\label{LS Sobolev}
\end{equation}
in $H^{1,-s}(\Bbb{R}^{3})$ $($which in fact belongs to $H^{2,-s}(\Bbb{R}%
^{3}))$ and this solution can be identified with the unique solution $\eta _{%
\mathrm{k}}$\ of the equation (\ref{LS_cont}) in $C_{\infty }(\Bbb{R}^{3})$%
.\bigskip 
\end{proposition}

\noindent \textbf{Proof.} Recall that any $V\in \mathrm{(I)}_{2}$ satisfies
condition (V.$m.\beta $) for $m=1,2$ and some $\beta >2.$ It follows that
the map $\kappa \mapsto G_{\kappa }V$ is a $\mathcal{B}_{\infty }(H^{m,-s})$%
-valued function, analytic in the upper half-plane and continuous in $%
\overline{\Bbb{C}}_{+}$. Then a variant of Fredholm theory (analogous to 
\cite{RS3}, Proposition on page 101) shows that there exists $%
(1+G_{k}V)^{-1}\in \mathcal{B}(H^{m,-s})$ for every $k\in [0,+\infty
)\setminus N_{m}$, where $N_{m}$ is a closed set of Lebesgue measure zero.
Moreover, $k$ belongs to $N_{m}$ if and only if there exists a non-zero
solution $\psi \in H^{m,-s}$ of the homogenous equation \ $(1+G_{k}V)\psi =0$%
.

\noindent From the fact that $H$ has no positive eigenvalues, it follows
that the previous homogenous equation can not have any non-zero solution in $%
H^{2,-s}$ if $k\neq 0$ (this follows from \cite{Agmon}, Th. 3.3; notice that 
$\psi \in H^{2,-s}$ implies $\hat{\psi}\in L_{\mathrm{loc}}^{1}$). \noindent
Then $N_{2}\subseteq \{0\}$.

\noindent Moreover, an argument similar to the proof of Prop. \ref{Prop
Equivalence} shows that any $\psi \in H^{1,-s}$ which solves the equation \ $%
(1+G_{k}V)\psi =0$ with $V\in \mathrm{(I)}_{2}$ really belongs to $H^{2,-s}$%
. Then $N_{1}\subseteq N_{2}\subseteq \{0\}$.

\noindent Since $g_{\mathrm{k}}\in H^{2,-s}$ for every $\mathrm{k}\in \Bbb{R}%
^{3}$, the unique solution of equation (\ref{LS Sobolev}) in $H^{2,-s}$ is
explicitly given by $\tilde{\eta}_{\mathrm{k}}=(1+G_{|\mathrm{k}|}V)^{-1}g_{%
\mathrm{k}}$ for every $\mathrm{k}\in \Bbb{R}^{3}\setminus \{0\}$. This is
also the unique solution of the equation in $H^{1,-s}$, since $%
H^{1,-s}\supseteq H^{2,-s}$ and $N_{1}\subseteq N_{2}$.

Now we prove that $\tilde{\eta}_{\mathrm{k}}\in H^{2,-s}$ can be identified
with $\eta _{\mathrm{k}}$ $\in C_{\infty }(\Bbb{R}^{3})$.

\noindent A straightforward argument, shows that the condition $f\in
H^{2,-s} $ is equivalent to the condition $\left\langle x\right\rangle
^{-s}f\in H^{2} $, where $\left\langle x\right\rangle ^{-s}$ represent the
multiplication times $(1+|\mathrm{x}|^{2})^{-\frac{s}{2}}$. \ Then, we
notice that the function 
\[
\tilde{\xi}_{\mathrm{k}}:=\left\langle x\right\rangle ^{-s}\tilde{\eta}_{%
\mathrm{k}}\in H^{2}(\Bbb{R}^{3})\subseteq C_{\infty }(\Bbb{R}^{3}) 
\]
satisfies the equation 
\begin{equation}
(1+\left\langle x\right\rangle ^{-s}G_{|\mathrm{k}|}V\left\langle
x\right\rangle ^{-s})\tilde{\xi}_{\mathrm{k}}=\left\langle x\right\rangle
^{-s}g_{\mathrm{k}}  \label{LS Aux}
\end{equation}
in $C_{\infty }$. On the other hand, \ \ also $\xi _{\mathrm{k}%
}:=\left\langle x\right\rangle ^{-s}\eta _{\mathrm{k}}$ satisfies equation (%
\ref{LS Aux}) in \ $C_{\infty }$, since $\eta _{\mathrm{k}}$ satisfies (\ref
{LS_cont}). An argument based on Fredholm theory shows that equation (\ref
{LS Aux}) has a unique solution in $C_{\infty }(\Bbb{R}^{3})$; then $\tilde{%
\xi}_{\mathrm{k}}=\xi _{\mathrm{k}}$ and this proves our claim.%
\endproof%
\bigskip

\noindent \textbf{Remark.} In the following we will not distinguish anymore
between $\tilde{\eta}_{\mathrm{k}}$ and $\eta _{\mathrm{k}}$. The pointwise
values of $\eta _{\mathrm{k}}$\bigskip will be denoted as $\eta _{\mathrm{k}%
}(\mathrm{y})\equiv \eta (\mathrm{y,k})$ for every $\mathrm{y}\in \Bbb{R}%
^{3} $.

We turn now to study the behavior of the generalized eigenfunctions in
presence of a zero-energy resonance. \ Although the following definition
could appear quite \textit{ad hoc}, we will show in the Appendix that it is
completely equivalent to the most common ones (see Prop. \ref{Prop
Equivalence}).

\begin{definition}
We say that there is a \textbf{zero-energy resonance} for the pair $%
(H,H_{0}) $ if \ there exists a $\psi \in H^{1,-\frac{1}{2}-0}(\Bbb{R}^{3})$
such that $\left( 1+G_{0}V\right) \psi =0$ but $\psi \notin L^{2}(\Bbb{R}%
^{3}).$\label{Def ZER}
\end{definition}

\noindent Such a $\psi $, if it exists, is unique up to a complex phase. It
will be called the \textbf{resonance function} and denoted with $\psi _{%
\mathrm{res}}$.

Roughly speaking, if the pair $(H,H_{0})$ admits a zero-energy resonance,
then the map $\kappa \mapsto (1+G_{\kappa }V)^{-1}$ has a singular behavior
as $\kappa \rightarrow 0$. \ This behavior is described by the following
theorem, which has been proved in \cite[Lemmas 4.2 and 4.4]{JK}.

\begin{theorem}[Jensen-Kato, 1979]
Assume that $V$ satisfies condition ($V.1.\beta $) with $\beta >7$. \ Let $s$
satisfy $7/2<s<\beta -7/2.$

Assume that there is a zero-energy resonance for the pair $(H,H_{0})$ and
that $0\notin \sigma _{\mathrm{p}}(H)$. \ Then for $\kappa \rightarrow
0,\kappa \in \Bbb{R}_{+}$, we have in $\mathcal{B}(H^{1,-s})$ the expansion 
\begin{equation}
(1+G_{\kappa }V)^{-1}=-\frac{i}{\kappa }\left\langle \cdot ,V\psi _{\mathrm{%
res}}\right\rangle \psi _{\mathrm{res}}+C_{0}+\kappa C_{1}+O(\kappa ^{2})
\label{JK claim}
\end{equation}
where $C_{0}$ and $C_{1}$ are explicitly computable operators in $\mathcal{B}%
(H^{1,-s})$.\label{Th JK}
\end{theorem}

An explicit expression for the operators $C_{0}$ and $C_{1}$ can be found in 
\cite[Lemma 4.3]{JK}. As \ pointed out in \cite[Remark 4.6]{JK}, by assuming
sufficiently large $\beta $ and $s$ it is possible to obtain an expansion of 
$(1+G_{\kappa }V)^{-1}$ to any order in $\kappa $. However, the actual
computation of coefficients becomes rather difficult.\label{Rem JKorder}\ 

\begin{remark}
If there is a zero-energy resonance for the pair $(H,H_{0})$ and moreover $%
0\in \sigma _{\mathrm{p}}(H)$ then expansion (\ref{JK claim}) is replaced by 
\begin{equation}
(1+G_{\kappa }V)^{-1}=-\frac{1}{\kappa ^{2}}P_{0}V-\frac{i}{\kappa }\left(
\left\langle \cdot ,V\psi _{\mathrm{res}}\right\rangle \psi _{\mathrm{res}%
}-P_{0}VCVP_{0}V\right) +\tilde{C}_{0}+O(1)  \label{JK claim2}
\end{equation}
where $P_{0}$ is the projector on the eigenspace relative to zero (naturally
extended to $H^{-1,-\frac{1}{2}+0}$ by using the fact that the
eigenfunctions belongs to $H^{1,\frac{1}{2}-0}$) and $C$ is the \
convolution operator with kernel $C(\mathrm{x},\mathrm{y})=\frac{1}{24\pi }|%
\mathrm{x}-\mathrm{y}|^{3}$.
\end{remark}

\begin{remark}
According to Jensen and Kato $($see \cite[Remark 6.7]{JK}$)$ the asymptotic
expansion (\ref{JK claim}) can be differentiated any number of times, in the
sense that for every $r\in \Bbb{N}$%
\begin{equation}
\frac{d^{r}}{d\kappa ^{r}}\left[ (1+G_{\kappa }V)^{-1}-\sum_{j=-1}^{n}\kappa
^{j}C_{j}\right] =o(\kappa ^{n-r}).  \label{JK diff}
\end{equation}
However, these asymptotic expansions require larger values of $s$\ and $%
\beta $ than for $r=0$.\label{Rem JKdiff} To fix notation, for every $n,r\in 
\Bbb{N}$ with $r<n$ there exists a real number $\bar{\beta}=\bar{\beta}(n,r)$
such that (\ref{JK claim2}) and (\ref{JK diff}) holds true, provided that $%
\beta >\bar{\beta}$ and $s$ satisfies $\frac{\bar{\beta}}{2}<s<\beta -\frac{%
\bar{\beta}}{2}$.
\end{remark}

It would be tempting to rephrase the previous result by saying that $%
(1+G_{\kappa }V)^{-1}$ has a ``pole'' in $\kappa =0$, but this term is
usually reserved for meromorphic functions (which are defined in an open
neighborhood of the point $\kappa =0$) while we are facing with a function
with an asymptotic expansion only on $\Omega =\{\kappa \in \Bbb{C}:\kappa
^{2}\in \overline{\Bbb{C}}_{+}\}$. In such a case we prefer to use the term 
\textbf{polar singularity}. The usual concept of complex analysis (simple
pole, residue,...) extends trivially to this case.

\noindent We emphasize that, when zero is a resonance but not an eigenvalue,
the expected polar singularity for $(1+G_{\kappa }V)^{-1}$ must be \emph{%
simple} \ and the residue corresponding to the pole is a \emph{rank-one
operator}, projecting on the subspace generated by the resonance function $%
\psi _{\mathrm{res}}\in H^{1,-\frac{1}{2}-0}(\Bbb{R}^{3}).\bigskip $

By recalling that the solution\ $\eta _{\mathrm{k}}$\ of the
Lippman-Schwinger equation (\ref{LS_cont}) is given by \ $\eta _{\mathrm{k}%
}=(1+G_{\kappa }V)^{-1}g_{\mathrm{k}}$ we can prove the following result.

\begin{proposition}
Assume that $V$ satisfies condition $($V.$1.\beta )$ for $\beta >\bar{\beta}%
\equiv \bar{\beta}(6,5)$ and that the pair $(H,H_{0})$ has a zero-energy
resonance. Fix $s$ so that $\frac{\bar{\beta}}{2}<s<\beta -\frac{\bar{\beta}%
}{2}$.

\noindent If $0\notin \sigma _{\mathrm{p}}(H),$ then the solution\ $\eta _{%
\mathrm{k}}$\ of the Lippman-Schwinger equation (\ref{LS Sobolev}) can be
decomposed as 
\begin{equation}
\eta (\mathrm{x},\mathrm{k})=\frac{r_{0}}{|\mathrm{k}|}\psi _{\mathrm{res}}(%
\mathrm{x})+\rho (\mathrm{x},\mathrm{k})\qquad  \label{Eta decompo}
\end{equation}
where $r_{0}=-i\left\langle g_{\mathrm{0}},V\psi _{\mathrm{res}%
}\right\rangle =i\left\langle V,\psi _{\mathrm{res}}\right\rangle $ and the
map $\mathrm{k}\mapsto \rho _{\mathrm{k}}$ from $\Bbb{R}^{3}\setminus \{0\}$
to $H^{1,-s}(\Bbb{R}^{3})$ is bounded (with all its derivatives until order $%
r=5$ at least) in a punctured neighborhood of the origin. \label%
{Prop sing eigen}

\noindent If $0\in \sigma _{\mathrm{p}}(H)$ and $\left\{ \psi _{j}\right\}
_{j=1}^{p}\subseteq H^{1,\frac{1}{2}-0}\subseteq L^{2}$ are the
corresponding eigenfunctions, then $\eta _{\mathrm{k}}$\ be decomposed as 
\[
\eta (\mathrm{x},\mathrm{k})=\frac{r_{0}}{|\mathrm{k}|}\psi _{\mathrm{res}}(%
\mathrm{x})+\sum_{j=1}^{p}\frac{r_{j}}{|\mathrm{k}|}\psi _{j}(\mathrm{x})+%
\bar{\rho}(\mathrm{x},\mathrm{k}) 
\]
where the map $\mathrm{k}\mapsto \bar{\rho}_{\mathrm{k}}$ $\ $has the same
properties stated above.\label{Prop reg}\bigskip
\end{proposition}

\noindent \textbf{Proof. \ }First consider the case\textbf{\ } $0\notin
\sigma _{\mathrm{p}}(H)$. \ From the expansion (\ref{JK claim}) (at higher
order) it follows that $\rho _{\mathrm{k}}$ is given by 
\[
\rho _{\mathrm{k}}=-\frac{i}{|\mathrm{k}|}\left\langle g_{\mathrm{k}}-g_{%
\mathrm{0}},V\psi _{\mathrm{res}}\right\rangle \psi _{\mathrm{res}}+C_{0}g_{%
\mathrm{k}}+|\mathrm{k}|C_{1}g_{\mathrm{k}}+\cdots +O(|\mathrm{k}|^{5}). 
\]
Then the claim follows from the differentiability of the map $\mathrm{k}%
\mapsto \left\langle g_{\mathrm{k}},V\psi _{\mathrm{res}}\right\rangle $
together with Remark \ref{Rem JKdiff}.

Now we turn to the case in which zero is an eigenvalue for $H$. By using the
expansion (\ref{JK claim2}) one gets that the coefficient of the
second-order pole is given by 
\[
P_{0}Vg_{\mathrm{0}}=\sum_{j=1}^{p}\left\langle \psi _{j},Vg_{\mathrm{0}%
}\right\rangle \psi _{j} 
\]
However, this term is identically zero since (we use the fact that $%
(1+G_{0}V)\psi _{j}=0$ since $\psi _{j}$ is an eigenfunction, see the
Appendix) one has 
\[
\left\langle \psi _{j},Vg_{\mathrm{0}}\right\rangle =\left\langle \psi
_{j},VG_{0}V\cdot 1\right\rangle =\left\langle G_{0}V\psi _{j},V\cdot
1\right\rangle =-\left\langle \psi _{j},V\cdot 1\right\rangle =0 
\]
where the last equality comes from the fact that $V\psi _{j}$ is orthogonal
(in the sense of dual pairing) to $1$ since $\psi _{j}$ is an \emph{%
eigenfunction} (see Lemma \ref{Lem Discrimen}).

\noindent Taking into account the fact that $P_{0}Vg_{\mathrm{0}}=0$ \ one
gets 
\[
\eta _{\mathrm{k}}=-\frac{1}{|\mathrm{k}|^{2}}\sum_{j=1}^{p}\left\langle
\psi _{j},V(g_{\mathrm{k}}-g_{\mathrm{0}})\right\rangle \psi _{j}-\frac{i}{|%
\mathrm{k}|}\left\langle V\psi _{\mathrm{res}},g_{\mathrm{k}}\right\rangle
\psi _{\mathrm{res}}+\tilde{C}_{0}g_{\mathrm{k}}+\cdots +O(|\mathrm{k}|^{5}) 
\]
and the claim follows as in the previous case. \bigskip 
\endproof%

The previous lemma gives relevant information about the behavior of the
generalized eigenfunctions in a neighborhood of the point $\mathrm{k=0}$. \
What about the behavior away from the origin in momentum space? \ The
regularity in $\mathrm{k}$ of the generalized eigenfunctions has been
studied in depth in \cite{TDMB}. The following result \ follows from the
proof of Th. 3.1 in the cited paper.

\begin{proposition}
\ Let be\ $V\in \mathrm{(I)}_{n}$ for some $n\geq 3$. Then \label%
{Prop Teufel}

\begin{enumerate}
\item  for every fixed $\mathrm{x}\in \Bbb{R}^{3}$, the function $\Phi _{\pm
}(\mathrm{x},\mathrm{\cdot })$ belongs to $C^{n-2}(\Bbb{R}^{3}\setminus
\{0\})$ and the partial derivatives $\partial _{k}^{\alpha }\Phi _{\pm }(%
\mathrm{x},\mathrm{k})$ $($for every multindex $\alpha \in \Bbb{N}^{3}$ with 
$|\alpha |\leq n-2)$ are continuous with respect to $\mathrm{x}\in \Bbb{R}%
^{3}$ and $\mathrm{k}\in \Bbb{R}^{3}\setminus \{0\}.$

\item  for every compact set $K\subseteq \Bbb{R}^{3}$ containing the origin 
\begin{equation}
\stackunder{\mathrm{k}\in \Bbb{R}^{3}\setminus K,\mathrm{x}\in \Bbb{R}^{3}}{%
\mathrm{Sup}}\left| \Phi _{\pm }(\mathrm{x},\mathrm{k})\right| \leq c_{K}
\label{Controllo1}
\end{equation}
\begin{equation}
\stackunder{\mathrm{k}\in \Bbb{R}^{3}\setminus K,\mathrm{x}\in \Bbb{R}^{3}}{%
\mathrm{Sup}}\left| \partial _{k}^{\alpha }\Phi _{\pm }(\mathrm{x},\mathrm{k}%
)\right| \leq c_{K,\alpha }(1+|x|)^{|\alpha |}  \label{Controllo2}
\end{equation}
for every $\alpha \in \Bbb{N}^{3}$ with $|\alpha |\leq n-2$.\bigskip
\end{enumerate}
\end{proposition}

\subsection{The flux-across-surfaces theorem}

We are now in position to study the FAS problem for an hamiltonian $%
H=H_{0}+V $ with a zero-energy resonance. We treat first the case in which
zero is not an eigenvalue.

First of all, let us focus on the properties of the asymptotic outgoing
state $\Psi _{\mathrm{out}}$ corresponding to a given $\Psi _{0}\in \mathcal{%
S}(\Bbb{R}^{3})$. From the eigenfunction expansion theorem we get

\begin{eqnarray}
\widehat{\Psi }_{\mathrm{out}}(\mathrm{k}) &=&\left( \mathcal{F}_{+}\Psi
_{0}\right) (\mathrm{k})=\int_{\Bbb{R}^{3}}\Phi _{+}(\mathrm{x},\mathrm{k}%
)^{*}\Psi _{0}(\mathrm{x})\ (2\pi )^{-\frac{3}{2}}dx  \nonumber \\
&=&\widehat{\Psi }_{0}(\mathrm{k})+\int_{\Bbb{R}^{3}}\eta (\mathrm{x},%
\mathrm{k})^{*}\Psi _{0}(\mathrm{x})\ (2\pi )^{-\frac{3}{2}}dx.
\end{eqnarray}

\noindent The local behavior of $\widehat{\Psi }_{\mathrm{out}}$ can be
analyzed using essentially Proposition \ref{Prop sing eigen}. Indeed, by
using the decomposition (\ref{Eta decompo}) we get

\begin{equation}
\widehat{\Psi }_{\mathrm{out}}(\mathrm{k})=\widehat{\Psi }_{0}(\mathrm{k})+%
\frac{r}{|\mathrm{k}|}+\int_{\Bbb{R}^{3}}\rho (\mathrm{x},\mathrm{k}%
)^{*}\Psi _{0}(\mathrm{x})\ dx  \label{Psi out dec}
\end{equation}
where $r=i\left\langle g_{\mathrm{0}},V\psi _{\mathrm{res}}\right\rangle
^{*}\left\langle \psi _{\mathrm{res}},\Psi _{0}\right\rangle $.

\noindent Decomposition (\ref{Psi out dec}) shows the typical \emph{singular
behavior} of the asymptotic outgoing state when the hamiltonian exhibits a
zero energy resonance, and Proposition \ref{Prop reg} assure that the
remaining part is \emph{regular} in some suitable sense.

On the contrary, we have little control on the asymptotic decrease of $%
\widehat{\Psi }_{\mathrm{out}}(\mathrm{k})$ as $|\mathrm{k}|\rightarrow
+\infty $. \ Indeed, the mapping properties of the wave operators, which
usually assure that $\Psi _{\mathrm{out}}$ inherits the properties of $\Psi
_{0},$ fail to hold when the hamiltonian has a zero-energy resonance (see 
\cite{Ya}). For this reason we are forced to assume \ \textit{a priori} a
suitable decrease of \ $\widehat{\Psi }_{\mathrm{out}}$ as $|\mathrm{k}%
|\rightarrow +\infty $. We will assume the following assumption (DA): $%
\widehat{\Psi }_{\mathrm{out}}\in C^{5}(\Bbb{R}^{3}\backslash \{0\})$ and
for every $m=0,...,5$ 
\begin{equation}
\left| \frac{\partial ^{m}}{\partial |\mathrm{k}|_{{}}^{m}}\widehat{\Psi }_{%
\mathrm{out}}(\mathrm{k})\right| \leq \frac{C_{m}}{|\mathrm{k}%
|^{3+\varepsilon +m}}\qquad \text{for\qquad }\left| \mathrm{k}\right| \geq
K_{m}  \tag{DA}  \label{Psi out decr}
\end{equation}
where $C_{m}$, $K_{m}$ and $\varepsilon $ are suitable positive constants.

\noindent Although this condition is stronger than what we proved in the
case of point interaction (see Lemma \ref{Lemma Out state}) the estimate (%
\ref{Psi out dec}) can be proved to hold for some solvable smooth potential.
For example, the \emph{Bargmann potential} 
\[
V_{b}(\mathrm{x})=-\frac{2b^{2}}{\cosh ^{2}(b|\mathrm{x}|)}\qquad (b>0) 
\]
admits a \ zero-energy resonance and, by using the explicit form of the
Lippman-Schwinger radial wavefunctions (see, for example, \cite{Barg})$^{(}$%
\footnote{%
It is well-known that, for a spherically simmetric potential, a zero-energy
resonance affects only the $s$-wave component of the scattering operator
(see, e.g. \cite{Ba}). Then one needs only the $s$-wave component of the
Lippman-Schwinger wavefunction.}$^{)}$ it is possible to prove that for any $%
\Psi _{0}\in \mathcal{S}(\Bbb{R}^{3})$ the corresponding $\widehat{\Psi }_{%
\mathrm{out}}$ decreases at infinity, with its derivatives, faster than the
inverse of any polynomial.

\begin{theorem}
Assume $V\in \mathrm{(I)}_{\infty }$ \ and that the hamiltonian $H=H_{0}+V$
\ has a zero-energy resonance or/and eigenvalue. \noindent Let be $\Psi
_{0}\in \mathcal{H}_{ac}(H)\cap \mathcal{S}(\Bbb{R}^{3})$ such that the
corresponding asymptotic outgoing state $\Psi _{\mathrm{out}}=W_{+}^{-1}\Psi
_{\mathrm{o}}$ satisfies the assumption (DA). \label{Th genFAS} \noindent
Then relation (FAS) holds, for every $T_{1}\in \Bbb{R}$.
\end{theorem}

\noindent \textbf{Remark.} We emphasize that the hypothesis $V\in \mathrm{(I)%
}_{\infty }$ has been assumed only for sake of simplicity. The proof works
assuming that $V\in \mathrm{(I)}_{n}$ for $n$ sufficiently large.\bigskip

\noindent We keep the notation as close as possible to the notation used in
Section 2 and in \cite{TDMB}.\bigskip

\noindent \textbf{Proof.} The first part of the proof \ follows closely the
proof of Theorem \ref{Th pointFAS}. Using the properties of $\mathcal{F}_{+}$
and (\ref{WOP relations}) we obtain 
\begin{eqnarray*}
\Psi _{t}(\mathrm{x}) &=&\int_{\Bbb{R}^{3}}e^{-i\mathrm{k}^{2}t}\widehat{%
\Psi }_{\mathrm{out}}(\mathrm{k})\Phi _{+}(\mathrm{x},\mathrm{k})\,(2\pi
)^{-3/2}dk \\
&=&\int_{\Bbb{R}^{3}}e^{-i\mathrm{k}^{2}t}\widehat{\Psi }_{\mathrm{out}}(%
\mathrm{k})e^{i\mathrm{k}\mathbf{\cdot }\mathrm{x}}\,(2\pi )^{-3/2}dk+\int_{%
\Bbb{R}^{3}}e^{-i\mathrm{k}^{2}t}\widehat{\Psi }_{\mathrm{out}}(\mathrm{k}%
)\eta (\mathrm{x},\mathrm{k})(2\pi )^{-3/2}dk \\
&\equiv &\alpha (\mathrm{x},t)+\beta (\mathrm{x},t)\,.
\end{eqnarray*}
\noindent As in the case of point interaction, the claim follows if one can
prove that 
\begin{equation}
\lim_{R\rightarrow \infty }\int_{T}^{+\infty }dt\int_{\Sigma _{R}}|\mathrm{j}%
_{1}(\mathrm{x},t)\mathbf{\cdot }\mathrm{n}|\,d\sigma =0\,  \label{j1 PS}
\end{equation}
where $\mathrm{j}_{1}\equiv \func{Im}(\alpha ^{*}\nabla \beta +\beta
^{*}\nabla \alpha +\beta ^{*}\nabla \alpha )$. In order to prove (\ref{j1 PS}%
) we need good estimates on $\alpha ,\beta $ and their gradients. \ We focus
first on the case in which zero is not an eigenvalue. \ \ 

\paragraph{Estimates on $\alpha $ and $\nabla \alpha $}

One decomposes $\alpha $ as $\alpha _{\mathrm{reg}}+\alpha _{\mathrm{sing}}$
where

\begin{eqnarray*}
\alpha _{\mathrm{reg}}(\mathrm{x},t) &=&\int_{\Bbb{R}^{3}}e^{i\mathrm{k}%
\mathbf{\cdot }\mathrm{x}}e^{-ik^{2}t}\,\left( \widehat{\Psi }_{\mathrm{out}%
}(\mathrm{k})-\frac{r}{|\mathrm{k}|}e^{-|\mathrm{k}|^{2}}\right) (2\pi
)^{-3/2}dk \\
\alpha _{\mathrm{sing}}(\mathrm{x},t) &=&\int_{\Bbb{R}^{3}}e^{i\mathrm{k}%
\mathbf{\cdot }\mathrm{x}}e^{-ik^{2}t}\frac{r}{|\mathrm{k}|}e^{-|\mathrm{k}%
|^{2}}\,(2\pi )^{-3/2}dk
\end{eqnarray*}
and from the properties of distributional Fourier transform we get 
\[
\nabla \alpha (\mathrm{x},t)=i\int_{\Bbb{R}^{3}}e^{i\mathrm{k}\mathbf{\cdot }%
\mathrm{x}}e^{-ik^{2}t}\mathrm{k}\widehat{\Psi }_{\mathrm{out}}(\mathrm{k}%
)\,(2\pi )^{-3/2}dk. 
\]
Decomposition (\ref{Psi out dec}) and Proposition \ref{Prop reg} show that
the functions 
\begin{eqnarray}
f_{1}(\mathrm{k}) &=&\widehat{\Psi }_{\mathrm{out}}(\mathrm{k})-\frac{r}{|%
\mathrm{k}|}e^{-|\mathrm{k}|^{2}}  \label{f1  PS} \\
f_{2}(\mathrm{k}) &=&\mathrm{k}\widehat{\Psi }_{\mathrm{out}}(\mathrm{k})
\label{f2  PS}
\end{eqnarray}
satisfy the condition (\ref{f hp1}). Moreover,assumption (DA) implies that
they satisfy also (\ref{f hp2}). So from Lemma \ref{FreeEvolution} it
follows that

\begin{equation}
\alpha _{\mathrm{reg}}=\frac{1}{t^{3/2}}\mathcal{O}_{1}\left( \frac{|\mathrm{%
x}|}{t}\right) \qquad \text{and\qquad }\nabla \alpha =\frac{1}{t^{3/2}}%
\mathcal{O}_{1}\left( \frac{|\mathrm{x}|}{t}\right)  \label{Alpha st1 PS}
\end{equation}
Finally, as in the proof of Theorem \ref{Th pointFAS} one proves that 
\begin{equation}
\alpha _{\mathrm{sing}}=\frac{1}{|\mathrm{x}|}\frac{1}{t^{1/2}}\mathcal{O}%
_{[-1,1]}\left( \frac{|\mathrm{x}|}{\sqrt{t}}\right) .  \label{Alpha st2 PS}
\end{equation}

\paragraph{Estimates on $\beta $ and $\nabla \beta .$}

We use the Lippman-Schwinger equation and, by Fubini's theorem, we get 
\begin{eqnarray*}
\beta (\mathrm{x},t)\, &=&\int_{\Bbb{R}^{3}}e^{-i\mathrm{k}^{2}t}\widehat{%
\Psi }_{\mathrm{out}}(\mathrm{k})\int_{\Bbb{R}^{3}}\frac{e^{-i|\mathrm{k}||%
\mathrm{x-y}|}}{|\mathrm{x-y}|}\ V(\mathrm{y})\ \Phi _{+}(\mathrm{y},\mathrm{%
k})(2\pi )^{-3/2}\ dy\ dk \\
&=&\int_{\Bbb{R}^{3}}\frac{V(\mathrm{y})}{|\mathrm{x-y}|}\int_{\Bbb{R}%
^{3}}e^{-i\left( \mathrm{k}^{2}t+|\mathrm{k}||\mathrm{x-y}|\right) }{}\ 
\widehat{\Psi }_{\mathrm{out}}(\mathrm{k})\ \Phi _{+}(\mathrm{y},\mathrm{k}%
)\ (2\pi )^{-3/2}\ dy\ dk \\
&\equiv &\int_{\Bbb{R}^{3}}\frac{V(\mathrm{y})}{|\mathrm{x-y}|}\Upsilon (%
\mathrm{x},\mathrm{y},t)\ dy.
\end{eqnarray*}
We extract from $\Upsilon $ the singular contributions. By standard
arguments we show that the function 
\begin{equation}
\tilde{\rho}(\mathrm{x},\mathrm{k})=\eta (\mathrm{x},\mathrm{k})-\frac{%
r_{0}^{{}}}{|\mathrm{k}|}e^{-|\mathrm{k}|^{2}}\psi _{\mathrm{res}}(\mathrm{x}%
)  \label{Ro tilde}
\end{equation}
has all the regularity properties claimed in Proposition \ref{Prop reg}. By
using definition (\ref{f1 PS}) and decomposition (\ref{Ro tilde}) we get 
\[
\Upsilon (\mathrm{x},\mathrm{y},t)=\int_{\Bbb{R}^{3}}e^{-i\left( \mathrm{k}%
^{2}t+|\mathrm{k}||\mathrm{x-y}|\right) }{}\left( ^{{}}\frac{r}{|\mathrm{k}|}%
e^{-|\mathrm{k}|^{2}}+f_{1}(\mathrm{k})\right) \left( e^{i\mathrm{k\cdot y}}+%
\frac{r_{0}}{|\mathrm{k}|}e^{-|\mathrm{k}|^{2}}\psi _{\mathrm{res}}(\mathrm{y%
})+\tilde{\rho}(\mathrm{y},\mathrm{k})\right) \tfrac{dk}{(2\pi )^{3/2}} 
\]
so we can decompose $\Upsilon $ as $\Upsilon _{\mathrm{sing},2}+\Upsilon _{%
\mathrm{sing},1}+\Upsilon _{\mathrm{reg}}$ where 
\begin{eqnarray*}
\Upsilon _{\mathrm{sing},2}(\mathrm{x},\mathrm{y},t) &=&\psi _{\mathrm{res}}(%
\mathrm{y})\int_{\Bbb{R}^{3}}e^{-i\left( \mathrm{k}^{2}t+|\mathrm{k}||%
\mathrm{x-y}|\right) }{}^{{}}e^{-2|\mathrm{k}|^{2}}\frac{rr_{0}}{|\mathrm{k}%
|^{2}}(2\pi )^{-3/2}\ dk, \\
\Upsilon _{\mathrm{sing},1}(\mathrm{x},\mathrm{y},t) &=&\psi _{\mathrm{res}}(%
\mathrm{y})\int_{\Bbb{R}^{3}}e^{-i\left( \mathrm{k}^{2}t+|\mathrm{k}||%
\mathrm{x-y}|\right) }{}^{{}}e^{-|\mathrm{k}|^{2}}f_{1}(\mathrm{k})\frac{%
r_{0}}{|\mathrm{k}|}(2\pi )^{-3/2}\ dk, \\
\Upsilon _{\mathrm{reg}}(\mathrm{x},\mathrm{y},t) &=&\int_{\Bbb{R}%
^{3}}e^{-i\left( \mathrm{k}^{2}t+|\mathrm{k}||\mathrm{x-y}|\right)
}{}^{{}}f_{1}(\mathrm{k})\left( e^{i\mathrm{k\cdot y}}+\tilde{\rho}(\mathrm{y%
},\mathrm{k})\right) (2\pi )^{-3/2}\ dk.
\end{eqnarray*}
An explicit gaussian integration gives 
\begin{equation}
\Upsilon _{\mathrm{sing},2}(\mathrm{x},\mathrm{y},t)=\frac{1}{\sqrt{t}}%
\mathcal{O}_{1}\left( \frac{|\mathrm{x-y}|}{\sqrt{t}}\right) \psi _{\mathrm{%
res}}(\mathrm{y}).  \label{Y st1}
\end{equation}

\noindent As for $\Upsilon _{\mathrm{sing},1}$, we perform explicit
integration of the singular part and we use a stationary phase method (see
Estimates on $\beta $ in Sec. 2) on the regular part getting 
\begin{equation}
\Upsilon _{\mathrm{sing},1}(\mathrm{x},\mathrm{y},t)=\frac{1}{t}\mathcal{O}%
_{2}\left( \frac{|\mathrm{x-y}|}{\sqrt{t}}\right) \psi _{\mathrm{res}}(%
\mathrm{y}).  \label{Y st2}
\end{equation}
>From (\ref{Y st1}) and (\ref{Y st2}) it follows then that the functions 
\[
\beta _{\mathrm{sing},j}(\mathrm{x},t)\equiv \int_{\Bbb{R}^{3}}\frac{V(%
\mathrm{y})}{|\mathrm{x-y}|}\Upsilon _{\mathrm{sing},j}(\mathrm{x},\mathrm{y}%
,t)\ dy\qquad (j=1,2) 
\]
satisfy the bound (\ref{Beta st1}).

The estimate on $\Upsilon _{\mathrm{reg}}$ is not so simple, since we have
only an indirect control on the behavior of $\tilde{\rho}$ . From 
\cite[Th. 3.1]{TDMB} (see Proposition \ref{Prop Teufel}) it follows that $%
\eta (\mathrm{y},\mathrm{\cdot })\in C^{1}(\Bbb{R}^{3}\setminus \{0\})$ for
every $\mathrm{y}\in \Bbb{R}^{3}$. \ We can then employ the stationary phase
technique of Section 2 in order to obtain 
\begin{equation}
\Upsilon _{\mathrm{reg}}(\mathrm{x},\mathrm{y},t)\leq \frac{C}{|\mathrm{x-y}%
|+t}\int_{\Bbb{R}^{3}}{}^{{}}\left| \frac{d}{dk}\left( \frac{1}{\chi
^{\prime }(k)}\ f_{1}(\mathrm{k})\left( e^{i\mathrm{k\cdot y}}+\tilde{\rho}(%
\mathrm{y},\mathrm{k})\right) k^{2}\right) \right| (2\pi )^{-3/2}\ dk\
d\omega  \label{Y reg expr}
\end{equation}
where we posed $\mathrm{k=}k\omega $ and $d\omega $ is the Lebesgue measure
on the sphere. Now we prove that 
\begin{equation}
\beta _{\mathrm{reg}}(\mathrm{x},t)\equiv \int_{\Bbb{R}^{3}}\frac{V(\mathrm{y%
})}{|\mathrm{x-y}|}\Upsilon _{\mathrm{reg}}(\mathrm{x},\mathrm{y},t)\ dy\leq 
\frac{C}{|\mathrm{x}|\left( |\mathrm{x}|+t\right) }.  \label{Beta st2 PS}
\end{equation}
By computing explicitly the derivative in (\ref{Y reg expr}) we get the
expression 
\[
\int_{\Bbb{R}^{3}}\frac{V(\mathrm{y})}{|\mathrm{x-y}|}\frac{C}{|\mathrm{x-y}%
|+t}\int_{\Bbb{R}^{3}}\xi _{1}(\mathrm{k})\tilde{\rho}(\mathrm{y},\mathrm{k}%
)+\xi _{2}(\mathrm{k})\frac{d\tilde{\rho}}{dk}(\mathrm{y},\mathrm{k})\ dk\
dy 
\]
where $\xi _{1}$ and $\xi _{2}$ are given explicitly by computing the
derivative. Now we observe that 
\begin{equation}
\Theta _{1}\equiv \int_{B_{1}(0)}\left( \xi _{1}(\mathrm{k})\tilde{\rho}_{%
\mathrm{k}}+\xi _{2}(\mathrm{k})\frac{d\tilde{\rho}_{\mathrm{k}}}{dk}\right)
\ dk\in H^{1,-s}(\Bbb{R}^{3})  \label{B integral}
\end{equation}
since $\xi _{1}$, $\xi _{2},$ $\tilde{\rho}_{k}$ and $\frac{d\tilde{\rho}}{dk%
}$ are bounded in every neighborhood of the origin (see Prop. \ref{Prop reg}%
). The condition $\Theta _{1}\in H^{1,-s}(\Bbb{R}^{3})$ it is sufficient to
prove that 
\[
\int_{\Bbb{R}^{3}}\frac{\left| V(\mathrm{y})\right| }{|\mathrm{x-y}|}\ \frac{%
\left| \Theta _{1}(\mathrm{y})\right| }{|\mathrm{x-y}|+t}dy\leq \frac{C}{|%
\mathrm{x}|\left( |\mathrm{x}|+t\right) }. 
\]

\noindent As for the remaining part, given by 
\[
\Theta _{2}(\mathrm{y})\equiv \int_{\Bbb{R}^{3}\setminus B_{1}(0)}\left( \xi
_{1}(\mathrm{k})\tilde{\rho}_{\mathrm{k}}+\xi _{2}(\mathrm{k})\frac{d\tilde{%
\rho}_{\mathrm{k}}}{dk}\right) \ dk, 
\]
we obtain that 
\[
\left| \Theta _{2}(\mathrm{y})\right| \leq \int_{\Bbb{R}^{3}\setminus
B_{1}(0)}\left| \xi _{1}(\mathrm{k})\right| \left\| \tilde{\rho}_{\mathrm{k}%
}\right\| _{\infty }+\left| \xi _{2}(\mathrm{k})\right| \left\| \frac{d%
\tilde{\rho}_{\mathrm{k}}}{dk}\right\| _{\infty }\ dk. 
\]
The decrease of $\xi _{j}$ \ $(j=1,2)$ and the uniform bounds on $\tilde{\rho%
}_{\mathrm{k}}$ and $\frac{d\tilde{\rho}_{\mathrm{k}}}{dk}$ are sufficient
to show that the last integral is finite. This proves (\ref{Beta st2 PS}).

\noindent As far as $\nabla \beta $ is concerned, a similar argument give
the estimate (\ref{Beta st3}).\bigskip

Finally, we use estimates on $\alpha $, $\nabla \alpha $, $\beta $ and $%
\nabla \beta $ in order to prove (\ref{j1 PS}). Notice that, in the present
case, the most singular part of \ $\beta ^{*}\nabla \beta $ is \emph{not real%
} as was in the case of point interaction. The most singular part (the only
one that cannot be controlled by using the previous estimates) is explicitly
given by 
\begin{eqnarray*}
\mathrm{j}_{\mathrm{cr}}(\mathrm{x},t) &=&\func{Im}\left( \int_{\Bbb{R}^{3}}%
\frac{V(\mathrm{y})}{|\mathrm{x-y}|}\Upsilon _{\mathrm{sing},2}(\mathrm{x},%
\mathrm{y},t)\ dy\right) ^{*}\left( -\frac{\mathrm{x}}{|\mathrm{x}|}\int_{%
\Bbb{R}^{3}}\frac{V(\mathrm{y}^{\prime })}{|\mathrm{x-y}^{\prime }|^{2}}%
\Upsilon _{\mathrm{sing},2}(\mathrm{x},\mathrm{y}^{\prime },t)\ dy^{\prime
}\right) \\
&=&C\frac{\mathrm{x}}{|\mathrm{x}|}\int_{\Bbb{R}^{3}}\int_{\Bbb{R}^{3}}\frac{%
V(\mathrm{y})\psi _{\mathrm{res}}(\mathrm{y})}{|\mathrm{x-y}|}\frac{V(%
\mathrm{y}^{\prime })\psi _{\mathrm{res}}(\mathrm{y}^{\prime })}{|\mathrm{x-y%
}^{\prime }|^{2}}\func{Im}\left( \varphi _{\mathrm{cr}}^{*}(\mathrm{x}-%
\mathrm{y},t)\varphi _{\mathrm{cr}}(\mathrm{x}-\mathrm{y}^{\prime
},t)\right) \ dy\ dy^{\prime }
\end{eqnarray*}
where we used the reality of $\psi _{\mathrm{res}}$ (see Sec. 4) and we
defined

\[
\varphi _{\mathrm{cr}}(\mathrm{x}-\mathrm{y},t)\equiv \int_{\Bbb{R}%
^{3}}e^{-i\left( \mathrm{k}^{2}t+|\mathrm{k}||\mathrm{x-y}|\right)
}{}^{{}}e^{-|\mathrm{k}|^{2}}\frac{1}{|\mathrm{k}|^{2}}\ dk=\frac{\sqrt{\pi }%
}{2}\frac{1}{\sqrt{1+it}}\left\{ e^{z^{2}}\mathrm{erfc}(z)\right\} _{z=\frac{%
i|\mathrm{x-y}|}{\sqrt{1+it}}}. 
\]
Now we observe that 
\[
\varphi _{\mathrm{cr}}(\mathrm{x}-\mathrm{y},t)=\frac{c}{\sqrt{t}}\left\{
e^{z^{2}}-e^{z^{2}}\mathrm{\func{erf}}(z)\right\} _{z=\frac{i|\mathrm{x-y}|}{%
\sqrt{1+it}}} 
\]
Since the second term has a first order zero in $z=0$ we can show that 
\[
\varphi _{\mathrm{cr}}(\mathrm{x}-\mathrm{y},t)=\frac{c}{\sqrt{t}}\left\{
e^{z^{2}}\right\} _{z=\frac{i|\mathrm{x-y}|}{\sqrt{1+it}}}+\frac{1}{\sqrt{t}}%
\mathcal{O}_{[-1,0]}\left( \frac{|\mathrm{x}|}{\sqrt{t}}\right) 
\]
The property $\mathcal{O}_{[-1,0]}$ of the second term can be now used to
improve the decrease in time of the corresponding terms, leading to the
usual vanishing argument. The only difficult term is then 
\[
\func{Im}(\exp (z^{2}+\left( z^{\prime }\right) ^{2}))=e^{-\frac{1}{1+t^{2}}%
\left( |\mathrm{x-y}^{\prime }|^{2}+|\mathrm{x-y}|^{2}\right) }\sin \left( 
\frac{t}{1+t^{2}}\left( |\mathrm{x-y}^{\prime }|^{2}-|\mathrm{x-y}%
|^{2}\right) \right) 
\]
and a convenient bound is given by 
\[
\left| \func{Im}(\exp (z^{2}+\left( z^{\prime }\right) ^{2}))\right| \leq 
\frac{C}{t^{1/4}}\left( |\mathrm{x-y}^{\prime }|^{2}+|\mathrm{x-y}%
|^{2}\right) ^{1/4}. 
\]
One proves that 
\[
\int_{\Bbb{R}^{3}}\int_{\Bbb{R}^{3}}\frac{\left( |\mathrm{x-y}^{\prime
}|^{2}+|\mathrm{x-y}|^{2}\right) ^{1/4}}{|\mathrm{x-y}||\mathrm{x-y}^{\prime
}|^{2}}V(\mathrm{y})\psi _{\mathrm{res}}(\mathrm{y})V(\mathrm{y}^{\prime
})\psi _{\mathrm{res}}(\mathrm{y}^{\prime })\ dy\ dy^{\prime }\leq \frac{C}{|%
\mathrm{x}|^{5/2}} 
\]
and then it follows that 
\[
\left| \mathrm{j}_{\mathrm{cr}}(\mathrm{x},t)\right| \leq \frac{C_{1}}{|%
\mathrm{x}|^{5/2}t^{5/4}}+\frac{C_{2}}{|\mathrm{x}|^{3}t}\mathcal{O}%
_{[-1,0]}\left( \frac{|\mathrm{x}|}{\sqrt{t}}\right) ^{2}. 
\]
The time decreasing of the first term and the property $\mathcal{O}_{[-1,0]}$
of the second term are sufficient to employ the dominated convergence
theorem and prove (\ref{j1 PS}). This concludes the proof of Theorem \ref{Th
genFAS} in the case in which zero is not an eigenvalue of $H$.

If zero is an eigenvalue, the proof follows a similar line, the main
ingredient being again Prop. \ref{Prop sing eigen} 
\endproof%

\newpage

\section{Appendix: some remarks on zero-energy resonances}

From a phenomenological point of view, the idea of \emph{quantum resonance}
is related to a ``bump'' of an observable quantity as a function of some
experimental parameter, like -- for example -- the scattering cross section
as a function of the energy of the incoming particles.$^{(}$\footnote{%
For a general overview of \ the relationship between the phenomenology and
the mathematical theory of resonance see \cite{AFS}\ or \cite{RS4}, Sec.
XII.6.\medskip}$^{)}$

\noindent The appearing of these ``bumps'' in the scattering cross-section
can be related, from a theoretical point of view, to some mathematical
properties of the pair $(H,H_{0})$, as, for example:

\begin{enumerate}
\item  the existence of almost-$L^{2}$ solutions of the stationary\ Schr\"{o}%
dinger equation

\item  the poles of a suitable meromorphic extension of the\emph{\ resolvent}

\item  the poles of the analytically continued \emph{scattering operator} in
the momentum decomposition.
\end{enumerate}

\noindent In many situations, as for example in the case of exponentially
decaying potentials, it can be shown that the previous properties are
equivalent.

\subsection{Resonances as complex poles}

Roughly speaking, the mathematical theory of resonances is based upon the
idea that, under suitable assumptions on the potential, there exists -- in
some suitable topology -- a convenient \emph{meromorphic continuation} of
the resolvent 
\[
R(\kappa )=(H-\kappa ^{2})^{-1}\qquad (\kappa \in \Bbb{C}_{+},\kappa ^{2}\in
\rho (H)) 
\]
to a region larger than $\Bbb{C}_{+}$. It is clear that the resolvent does
not admit a meromorphic continuation \ as a $\mathcal{B}(L^{2})$-valued
function, since for $\kappa \in \Bbb{R}$ we have $\kappa ^{2}\in \sigma (H)$%
. However, one introduces suitable Banach spaces $X$, $Y$ (sometimes with
the additional conditions $X\subseteq L^{2}$ and $H^{2}\subseteq Y$) such
that the resolvent admits a meromorphic continuation as a $\mathcal{B}(X,Y)$%
-valued function.$^{(}$\footnote{%
The choice of the spaces $X,Y$ $\ $-- although sometime suggested by the
properties of the potential -- is not natural and there is no \emph{a priori}
warranty that the position of the poles of the meromorphic continuation do
not depend on this choice. This ambiguity leads to the problem of the \emph{%
equivalence of definitions} of resonances (see Prop. \ref{Prop Equivalence}
for zero-energy resonances).\medskip}$^{)}$

\noindent One notices that, under general assumptions on the potential, the
resolvent $R(\kappa )\in \mathcal{B}(L^{2})$, satisfy the \emph{resolvent
equations }

\noindent 
\begin{equation}
R(\kappa )=(1+R_{0}(\kappa )V)^{-1}R_{0}(\kappa )  \label{RE1}
\end{equation}
\begin{equation}
R(\kappa )=R_{0}(\kappa )(1+VR_{0}(\kappa ))^{-1}  \label{RE2}
\end{equation}

\noindent and is meromorphic in $\Bbb{C}_{+}$ with poles at the discrete
eigenvalues of $H$. \ Decompositions (\ref{RE1}) and (\ref{RE2})\ suggest
defining the meromorphic continuation of the resolvent by an analytic
continuation of the \emph{free} resolvent and the previous factorization. $%
^{(}$\footnote{%
Sometime it is convenient to notice that, under the assumption that $V$
belongs to the Rollnik class, the resolvent can be expressed as (see \cite
{RS3}, Chapter XI, Problem 61 or \cite{Ge}) 
\[
R(\kappa )=R_{0}(\kappa )+\left( R_{0}(\kappa )|V|^{1/2}\right) \left(
1+K_{\kappa }\right) ^{-1}\left( V^{1/2}R_{0}(\kappa )\right) 
\]
where 
\[
K_{\kappa }:=V^{1/2}R_{0}(\kappa )|V|^{1/2}\qquad (\kappa \in \Bbb{C}_{+}). 
\]
\par
\noindent This symmetric decomposition can be useful in order to define a
convenient meromorphic extension of $R(\kappa )-R_{0}(\kappa )$ to a region $%
\Omega $ larger than $\Bbb{C}_{+}$.\label{Note Simon}}$^{)}\bigskip $

In order to illustrate this method, we focus first on the class of \emph{%
exponentially decaying} potentials and we follow an exposition dual to the
one given by \cite{Ba}. \ \ 

\ To be precise, we require that 
\begin{equation}
V=e^{-ar}We^{-ar}\qquad (r=|x|)  \label{Exp decay}
\end{equation}
where $W$ is an $H_{0}$-compact, symmetric operator. Under this assumption
the resolvent equations (\ref{RE1}) and (\ref{RE2}) hold true for every $%
\kappa \in \Bbb{C}_{+}$. \ It is convenient to define the spaces (for $%
\sigma \geq 0$) 
\[
H_{\sigma }^{{}}(\Bbb{R}^{d}):=\left\{ u\in \mathcal{S}^{\prime }(\Bbb{R}%
^{d}):\left\| e^{\sigma |x|}u\right\| _{L^{2}}<+\infty \right\} 
\]
\[
H_{\pm \sigma }^{2}(\Bbb{R}^{d}):=\left\{ u\in \mathcal{S}^{\prime }(\Bbb{R}%
^{d}):\left\| e^{\pm \sigma |x|}u\right\| _{H^{2}}<+\infty \right\} 
\]
since the free resolvent, regarded as a $\mathcal{B}(H_{\sigma
}^{{}},H_{-\sigma }^{2})$-valued function, admits a meromorphic continuation 
$\widetilde{R}_{0}(\kappa )$, to the region 
\[
\Omega _{\sigma }=\left\{ \kappa \in \Bbb{C}:\func{Im}(\kappa )>-\sigma
\right\} . 
\]
Condition (\ref{Exp decay}) implies that $V\in \mathcal{B}_{\infty
}(H_{-a}^{2},H_{a}^{{}}),$ where $a$ is given in (\ref{Exp decay}), so that
the map $\kappa \mapsto $ $1+\widetilde{R}_{0}(\kappa )V$ is a compact
operator valued analytic function on $\Omega _{a}$. Then the Fredholm theory
implies that the r.h.s of (\ref{RE1}) is a meromorphic function on $\Omega
_{a}$(with a discrete set of singularities $\Sigma $) that can be seen as a
meromorphic continuation of the resolvent. \noindent Moreover a point $%
\kappa _{0}\in \Omega _{a}$ belongs to $\Sigma $ if and only if the equation 
\begin{equation}
(1+\tilde{R}_{0}(\kappa _{0})V)\psi =0  \label{Ikebe}
\end{equation}
admits a non trivial solution $\psi \in H_{-a}^{2}$. \medskip

It is possible to prove that a point $\kappa _{0}\in \Sigma \ \cap \ \Bbb{C}%
_{+}$ must lie on the imaginary axis and corresponds to a discrete negative
eigenvalue of $H$ with eigenfunction $\psi $. \ \noindent A point $\kappa
_{0}\in \Sigma \ \cap \ $ $\left( \Bbb{R}\setminus \{0\}\right) $
corresponds to a positive eigenvalue of $H$ (embedded in the continuous
spectrum). \noindent So we are lead to the following definition.

\begin{definition}
A singular point $\kappa _{0}\in \Sigma $ in the lower half-plane $\Bbb{C}%
_{-}$ is called a \emph{\ }\textbf{resonance point} for the pair $(H,H_{0})$%
. A non trivial solution $\psi \in H_{-a}^{2}$ of the equation (\ref{Ikebe})
is called the \textbf{resonance function} corresponding to $\kappa _{0}$.
\end{definition}

\noindent We emphasize that the concept of resonance is related to the pair $%
(H,H_{0})$ and not only to the operator $H$, since decompositions (\ref{RE1}%
) and (\ref{RE2}) are based upon the decomposition of $H$ as $H_{0}+V$.

A singularity of the resolvent in the point $\kappa _{0}=$ $0$ corresponds
to the fact that the equation 
\begin{equation}
\left( 1+\tilde{R}_{0}(0)V\right) \psi =0  \label{ZRes eq}
\end{equation}
has non trivial solutions in $H_{-a}^{2}$. If there exists a solution $\psi
\in H_{-a}^{2}\cap L^{2}$ then zero is an eigenvalue for $H$. If there
exists a solution $\psi \in H_{-a}^{2}$ such that $\psi \notin L^{2}$ then
we say that there is a \emph{zero-energy resonance} and $\psi $ is called 
\emph{resonance function}. Notice that $\kappa _{0}=0$ can be both (and
simultaneously) an eigenvalue for $H$ and a resonance for the pair $%
(H,H_{0}) $ (see Lemma \ref{Lem Discrimen}).

For exponentially decaying potentials, it is possible to find \ a
relationship between to poles of the resolvent and the poles of the
analytically continued scattering matrix and then a connection with
observable quantities is possible (see again \cite{Ba} for details).

\subsection{General theory of zero-energy resonances}

In the previous analysis, we introduced the \emph{exponentially weighted}
Sobolev spaces $H_{\pm \sigma }^{m}$ in order to obtain an \emph{analytic}
continuation of the free resolvent to the open region $\Omega _{a}\supseteq 
\Bbb{C}_{+}$. This forced us to consider exponentially decreasing
potentials, to make $G_{0}V$ a well-defined compact operator on $H_{-a}^{2}$.

However, we are only interested in \emph{zero-energy resonances}, since only
the poles of the resolvent on the \emph{real axis} (in momentum complex
plane) can affect the behavior of the Lippman-Schwinger generalized
eigenfunctions. \ To define zero-energy resonances, we need only to \emph{%
continuously} extend the free resolvent to the closed upper half-plane, and
this extension is provided -- in the weighted Sobolev space topology -- by
Lemma \ref{Lemma Ext res}. Then formula (\ref{ZRes eq}) suggests to define
-- for a wide class of potential -- a zero energy resonance function as an
element of the null space of $1+G_{0}V\in \mathcal{B}(H^{2,-s})$ which does
not belong to $L^{2}$. The following analysis make this idea precise and
gives the connection with other usual approaches.

Assume condition (V.$1.\beta $) for $\beta >2.$ Then Lemma \ref{Lemma Ext
res} assures that $G_{0}V\in $ $\mathcal{B}_{\infty }(H^{1,-s})$ and $%
VG_{0}\in $ $\mathcal{B}_{\infty }(H^{-1,s})$, where $s>\frac{1}{2}$ and $%
\mathcal{B}_{\infty }$ denotes the space of compact operators. We denote
with $\frak{M}_{s}$ the kernel of \ $1+G_{0}V$ in $H^{1,-s}$ and with $\frak{%
N}_{s}$ the kernel of \ $1+VG_{0}$ in $H^{-1,s}$. These spaces may depend on 
$s$, but $\frak{M}_{s}$ is monotone increasing and $\frak{N}_{s}$ is
monotone decreasing in $s$ and -- by duality -- we have $\dim \frak{M}%
_{s}=\dim \frak{N}_{s}=d<+\infty $. Thus they are independent of $s$ and we
denote them simply as $\frak{M}$ and $\frak{N}$; moreover 
\[
\frak{M}\subseteq H^{1,-\frac{1}{2}-0}\qquad \frak{N}\subseteq H^{-1,\beta -%
\frac{1}{2}-0} 
\]

We summarize the properties of this spaces, proved by Jensen and Kato in 
\cite{JK}, in the following lemma.

\begin{lemma}
Fix $s\in (\frac{1}{2},\frac{3}{2}]$. Then:

\begin{enumerate}
\item  the kernel of $-\Delta +V$ in $H^{1,-s}$ coincides with $\frak{M}$

\item  both $-\Delta $ and $V$ are injective from $\frak{M}$ onto $\frak{N}$
and $G_{0}$ is injective from $\frak{N}$ onto $\frak{M}$ with inverse $%
-\Delta $. \ $^{(}$\footnote{%
In other words, while $-\Delta $ has only a \emph{right} inverse $G_{0}$ on $%
\mathcal{S}^{\prime }(\Bbb{R}^{d})$, the restriction $\left. -\Delta \right|
_{\frak{M}}$ has a \emph{bilateral} inverse $\left. G_{0}\right| _{\frak{N}}$%
}$^{)}$

\item  the eigenspace $P_{0}\mathcal{H}$ relative to the self-adjoint
operator $H$ in $L^{2}(\Bbb{R}^{3})$ is included in $\frak{M}\cap H^{1,\frac{%
1}{2}-0}$ with $\dim (\frak{M}$ $/P_{0}\mathcal{H})\leq 1$. Moreover, $\psi
\in \frak{M}$ \ belongs to $L^{2}$ if and only if $\left\langle \psi
,V1\right\rangle =0$. \label{Lem Discrimen}
\end{enumerate}
\end{lemma}

We notice that Lemma \ref{Lem Discrimen}.iii gives us a criterion to
distinguish in $\frak{M}$ $\ $a subspace (with codimension $\leq 1$)
corresponding to the eigenfunctions of the operator $H=H_{0}+V=\left.
-\Delta +V\right| _{L^{2}}$. We would like to associate the (possible)
complementary subspace to a zero-energy resonance. However, before
formulating a precise definition, we relate this approach to other possible
definitions of zero energy-resonances, in particular to the ones used in
Sec. 3 (in particular \cite{Ikebe}, \cite{TDMB}, \cite{JK}, \cite{Agmon}). A
connection with the approach outlined in Note \ref{Note Simon} is possible
too.

\begin{proposition}
Assume $V\in \mathrm{(I)}_{3}$. Then the following propositions are
equivalent: \label{Prop Equivalence}

\begin{enumerate}
\item  $\mathbf{(}$\textbf{standard definition}$)$ there exists a \emph{%
distributional} solution $\psi _{1}\in L_{\mathrm{loc}}^{2}(\Bbb{R}^{3})$ of
the stationary Schr\"{o}dinger equation $\left( -\Delta +V\right) \psi _{1}=0
$ \noindent such that $(1+|\cdot |^{2})^{-\frac{\gamma }{2}}\psi _{1}\in
L_{{}}^{2}(\Bbb{R}^{3})$ for every $\gamma >1/2$ but not for $\gamma =0$.

\item  $\mathbf{(}$\textbf{Agmon definition}$)$ there exists a $\psi _{2}\in
H^{2,-\gamma }\cap C_{\infty }$ $($for every $\gamma >\frac{1}{2})$ such
that $\left( 1+G_{0}V\right) \psi _{2}=0$ but $\psi _{2}\notin L^{2}$.

\item  $\mathbf{(}$\textbf{Jensen-Kato definition}$)$ there exists a $\psi
_{3}\in H^{1,-\gamma }$ $($for every $\gamma >\frac{1}{2})$ such that $%
\left( 1+G_{0}V\right) \psi _{3}=0$ but $\psi _{3}\notin L^{2}$.

\item  $\mathbf{(}$\textbf{dual Jensen-Kato definition}$)$ there exists a $%
\phi \in H^{-1,\gamma }$ $($for every $\gamma >\frac{1}{2})$ such that $%
\left( 1+VG_{0}\right) \phi =0$ but $G_{0}\phi \equiv \psi _{4}\notin L^{2}$.

\item  $\mathbf{(}$\textbf{Ikebe definition}$)$ there exists a $\psi _{5}\in
C_{\infty }(\Bbb{R}^{3})$ such that $\left( 1+G_{0}V\right) \psi _{5}=0$ but 
$\psi _{5}\notin L^{2}$.
\end{enumerate}

\noindent In addition, all the functions $\psi _{i}$ $(i=1,\ldots ,5)$ are
the same element of \ $\mathcal{S}^{\prime }(\Bbb{R}^{3})$ that will be
denoted by $\psi _{\mathrm{res}}$. Moreover, if the potential satisfies (\ref
{Exp decay}) (and is a multiplication operator) then $\psi _{\mathrm{res}%
}\in H_{-a}^{2}(\Bbb{R}^{3})$ and each one of the previous propositions
implies that the meromorphic continuation of the resolvent, seen as a $%
\mathcal{B}(H_{a}^{{}},H_{-a}^{2})$-valued function, has a complex pole in $%
\kappa =0$.
\end{proposition}

\begin{definition}[Zero-energy resonances]
We say that there is a zero-energy resonance for the pair $(H,H_{0})$ if \
one of the conditions of Proposition \ref{Prop Equivalence} holds true.
\end{definition}

\noindent \textbf{Remark.}\ From the fact that $-\Delta G_{0}=1$ it follows
straightforwardly that if $\psi \in \mathcal{S}^{\prime }(\Bbb{R}^{3})$
satisfy the equation

\begin{equation}
\left( 1+G_{0}V\right) \psi =0\text{ }  \label{Ikebe equation}
\end{equation}
then it satisfies also the equation 
\begin{equation}
-\Delta \psi +V\psi =0  \label{Sch  equation}
\end{equation}
provided that $V\psi $ is a well-defined distribution. $^{(}$\footnote{%
This condition is satisfied for our class of potentials if $\psi \in L_{%
\mathrm{loc}}^{2}$}$^{)}$ The converse statement is not true in general,
since a \emph{left} inverse of $-\Delta $ in $\mathcal{S}^{\prime }$ \ does
not exist.\bigskip

\noindent \textbf{Proof of proposition. }We will show that (i) $\Rightarrow $%
(ii) $\Rightarrow $((iii) or (v))$\Rightarrow $(i). \ Moreover, (iii)$%
\Leftrightarrow $(iv) as a consequence of Lemma \ref{Lem Discrimen}.ii .

\textbf{(i) }$\Rightarrow $\textbf{(ii).} Suppose \ that $\psi _{1}\equiv
\psi \in L_{\mathrm{loc}}^{2}(\Bbb{R}^{3})$ satisfy the Schr\"{o}dinger
equation in distributional sense. From the fact that the kernel of $-\Delta $
is closed in \ $\mathcal{S}^{\prime }(\Bbb{R}^{3})$\ it follows, by a
decomposition argument, that the general solution of equation (\ref{Sch
equation}) must be in the form 
\begin{equation}
\psi =G_{0}V\psi +\varphi  \label{General solution}
\end{equation}
where $\varphi $ is an harmonic distribution, i.e. $\Delta \varphi =0$. \ We
are going to show that from our assumptions it follows that \ $\varphi =0$.

Indeed $\varphi \in L_{-\gamma }^{2}(\Bbb{R}^{3})$ (in the following we
understand $\gamma \in (\frac{1}{2},+\infty )$). This follows from (\ref
{General solution}) by noticing that $\psi \in L_{-\gamma }^{2}(\Bbb{R}^{3})$
by hypothesis $^{(}$\footnote{%
Recall that $f\in L_{s}^{2}$ if and only if \ $(1+|\cdot |^{2})^{\frac{s}{2}%
}f\in L^{2}$ .}$^{)}$ and that $G_{0}V\psi $ also belongs to $L_{-\gamma
}^{2}(\Bbb{R}^{3})$, as can be proved by using the explicit form of $G_{0}$
as convolution operator and Jensen inequality.

In particular $\varphi $ is an harmonic distribution that belongs to $L_{%
\mathrm{loc}}^{1}$,\ and then it is (identifiable with) a smooth function in 
$C^{\infty }(\Bbb{R}^{3})$ (see \cite{LiebLoss}, Th. 9.3).

Moreover, $\varphi $ must be \emph{bounded}. In order to see this, suppose
that $\varphi $ is unbounded. Then it is possible to find a sequence $%
\left\{ \mathrm{x}_{n}\right\} _{n\in \Bbb{N}}$ such that \ $\left| \varphi (%
\mathrm{x}_{n})\right| \geq n$. The continuity of $\varphi $ implies that
this sequence cannot have accumulation points. Then it is possible to
extract a subsequence $\left\{ \mathrm{x}_{n_{j}}\equiv \mathrm{w}%
_{j}\right\} _{j\in \Bbb{N}}$ such that:

\begin{enumerate}
\item  $\left| \varphi (\mathrm{w}_{j})\right| \geq n_{j}\geq j$ \ and $\ |%
\mathrm{w}_{j}|\geq 2^{j}$

\item  if $\delta _{j}=\frac{1}{2}|\mathrm{w}_{j}|$ \ then \ $B_{\delta
_{j}}(\mathrm{w}_{j})\cap \ B_{\delta _{l}}(\mathrm{w}_{l})=\emptyset $ if $%
j\neq l$.
\end{enumerate}

\noindent By using the disjointedness of the balls \ $B_{j}\equiv B_{\delta
_{j}}(\mathrm{w}_{j})$ we get that 
\begin{eqnarray*}
\left\| \varphi \right\| _{L_{-\gamma }^{2}} &=&\int_{\Bbb{R}^{3}}\left|
\varphi (\mathrm{y})\right| ^{2}(1+|\mathrm{y}|^{2})^{-\gamma }\ d\mathrm{y}%
\geq \\
&\geq &\sum_{j\in \Bbb{N}}\int_{B_{j}}\left| \varphi (\mathrm{y})\right|
^{2}(1+|\mathrm{y}|^{2})^{-\gamma }\ d\mathrm{y}\geq \\
&\geq &\sum_{j\in \Bbb{N}}\stackunder{\mathrm{y}\in B_{j}}{\inf }\left\{ (1+|%
\mathrm{y}|^{2})^{-\gamma }\right\} \int_{B_{j}}\left| \varphi (\mathrm{y}%
)\right| ^{2}\ d\mathrm{y}
\end{eqnarray*}
>From the fact that $\Delta \varphi =0$ it follows that $\left| \varphi
\right| ^{2}$ is a subharmonic function, i.e. $\Delta \left| \varphi \right|
^{2}\geq 0$. Then we can apply the mean value inequality getting 
\begin{eqnarray*}
\left\| \varphi \right\| _{L_{-\gamma }^{2}} &\geq &\sum_{j\in \Bbb{N}}%
\stackunder{\mathrm{y}\in B_{j}}{\inf }\left\{ (1+|\mathrm{y}|^{2})^{-\gamma
}\right\} \frac{4}{3}\pi \delta _{j}^{3}\left| \varphi (\mathrm{w}%
_{j})\right| ^{2} \\
\ &\geq &\sum_{j\in \Bbb{N}}C\frac{\delta _{j}^{3}\quad j^{2}}{(1+(|\mathrm{w%
}_{j}|+\delta _{j})^{2})^{\gamma }}
\end{eqnarray*}
By using the claimed properties of the sequence $\left\{ \mathrm{w}%
_{j}\right\} _{j\in \Bbb{N}}$ one gets that the previous series diverges,
against the fact that $\varphi \in L_{-\gamma }^{2}(\Bbb{R}^{3})$. Then $%
\varphi $ is bounded. \ \ 

By Liouville's theorem, a bounded harmonic function on $\Bbb{R}^{d}$ is
constant and a constant function that belongs to $L_{-\gamma }^{2}(\Bbb{R}%
^{3})$ (for every $\gamma >\frac{1}{2}$) must be identically zero. Then $%
\varphi =0$.

This proves that $\psi $ satisfies equation (\ref{Ikebe equation}) in
distributional sense. It remains to be shown that $\psi \in H^{2,-\gamma
}\cap C_{\infty }$.

Since $\psi =G_{0}V\psi $, an iterative or ``bootstrap'' argument based upon
the smoothing properties of the \ operator $G_{0}$ (see, for example, \cite
{LiebLoss} Sec. 10.2) shows that $\psi \in C^{0,\alpha }(\Bbb{R}^{3})$, the
space of uniformly H\"{o}lder-continuous functions of order $\alpha $, for
every $\alpha <1/2$.\ 

Moreover, $\psi \in L^{\infty }(\Bbb{R}^{3})$. \ Recall that the hypothesis $%
V\in (\mathrm{I})_{n}$ implies that there exist $R>0$ and $\varepsilon >0$
such that $V$ is continuous in $\Bbb{R}^{3}\setminus B_{R}$ and 
\begin{equation}
\left| V(\mathrm{y})\right| \leq \frac{C}{|\mathrm{y}|^{n+\varepsilon }}%
\qquad \text{for }|\mathrm{y}|>R.  \label{Hp}
\end{equation}
Choose $\gamma _{0}=\frac{1}{2}+\frac{\varepsilon }{2}$ (with $\varepsilon $
from (\ref{Hp})). \noindent Fix a compact set $K$ such that $K\supseteq
B_{2R}$. \ The continuity of $\psi $ implies that it is bounded over the
compact set $K$. \ To show the boundness on $\Bbb{R}^{3}\setminus K,$ we use
the explicit form of the distributional kernel of $G_{0}$ getting

\begin{eqnarray*}
\left| \psi (\mathrm{x})\right| &\leq &\int_{\Bbb{R}^{3}\setminus B_{1}(%
\mathrm{x})}\frac{1}{|\mathrm{x-y}|}\left| V(\mathrm{y})\psi (\mathrm{y}%
)\right| \ dy+\int_{B_{1}(\mathrm{x})}\frac{1}{|\mathrm{x-y}|}\left| V(%
\mathrm{y})\psi (\mathrm{y})\right| \ dy \\
&\equiv &\psi _{1}(\mathrm{x})+\psi _{2}(\mathrm{x})
\end{eqnarray*}

\noindent The first term can be easily bounded, by noticing that 
\[
\left| \psi _{1}(\mathrm{x})\right| \leq \int_{\Bbb{R}^{3}}\left| V(\mathrm{y%
})\right| \left| \psi (\mathrm{y})\right| \ dy<+\infty 
\]
where the last inequality follows from the fact that $V\psi \in L_{\mathrm{%
loc}}^{1}$ \ and that for $V\in (\mathrm{I})_{2}$ one has 
\begin{equation}
\int_{\Bbb{R}^{3}\setminus B_{2R}(0)}\left| V(\mathrm{y})\right| \left| \psi
(\mathrm{y})\right| \ dy\leq \left( \int_{\Bbb{R}^{3}\setminus
B_{2R}(0)}\left| V(\mathrm{y})\right| ^{2}(1+|\mathrm{y}|)^{2\gamma
_{0}}dy\right) ^{1/2}\left\| \psi \right\| _{L_{-\gamma _{0}}^{2}}<+\infty .
\label{Bound Psi1}
\end{equation}

\noindent As for the second term, by a simple change of variables we get 
\begin{eqnarray}
\left| \psi _{2}(\mathrm{x})\right| &\leq &\int_{B_{1}(\mathrm{x})}\frac{1}{|%
\mathrm{x-y}|}\left| V(\mathrm{y})\psi (\mathrm{y})\right| \ dy  \nonumber \\
&\leq &\int_{B_{1}(\mathrm{0})}\frac{1}{|\mathrm{w}|}\left| V(\mathrm{w+x}%
)\psi (\mathrm{w+x})\right| \ dw  \nonumber \\
&\leq &\stackunder{\mathrm{y\in }B_{1}(\mathrm{x})}{\mathrm{Sup}}\left\{
\left| V(\mathrm{y})\right| (1+|\mathrm{y}|)^{\gamma }\right\} \int_{B_{1}(%
\mathrm{0})}\frac{1}{|\mathrm{w}|}(1+|\mathrm{w+x}|)^{-\gamma }\left| \psi (%
\mathrm{w+x})\right| \ dw  \nonumber \\
&\leq &\stackunder{\mathrm{y\in }B_{1}(\mathrm{x})}{\mathrm{Sup}}\left\{
\left| V(\mathrm{y})\right| (1+|\mathrm{y}|)^{\gamma }\right\} \left(
\int_{B_{1}(\mathrm{0})}\frac{1}{|\mathrm{w}|^{2}}\ dw\right) ^{1/2}\left\|
\psi \right\| _{L_{-\gamma }^{2}}  \label{Bound Psi2}
\end{eqnarray}
where in the last inequality we applied the Schwartz theorem.

\noindent For every $\mathrm{y\in }B_{1}(\mathrm{x})$ we have $\frac{|%
\mathrm{x}|}{2}<|\mathrm{y}|<\frac{3}{2}|\mathrm{x}|$ (without loss of
generality we can assume $R>1$, so that the previous condition holds true).
Recalling that $|\mathrm{x}|>2R$ and using the decreasing condition (\ref{Hp}%
) we obtain 
\begin{equation}
\stackunder{\mathrm{y\in }B_{1}(\mathrm{x})}{\mathrm{Sup}}\left\{ \left| V(%
\mathrm{y})\right| (1+|\mathrm{y}|)^{\gamma }\right\} \leq C^{\prime }\frac{%
(1+|\mathrm{x}|)^{\gamma }}{|\mathrm{x}|^{2+\varepsilon }}<C_{R}\qquad
\label{Bound C}
\end{equation}
where the last bound follows by choosing $\gamma \in (\frac{1}{2},2)$.
\noindent From estimates (\ref{Bound Psi1}), (\ref{Bound Psi2}) \ and (\ref
{Bound C}) it follows our claim $\psi \in L^{\infty }(\Bbb{R}^{3})$.

It has been proved by Ikebe (see \cite{Ikebe}, \ Lemma 3.1) that a bounded
and continuous solution of the equation (\ref{Ikebe equation}) with $V\in 
\mathrm{(I)}_{2}$ must vanish at infinity. Then $\psi \in $ $C_{\infty }(%
\Bbb{R}^{3})$.

Finally, we notice that $\Delta \psi =V\psi \in L_{\mathrm{loc}}^{2}$ and
hence, by Sobolev inequalities, we get $\psi \in H_{\mathrm{loc}}^{2}$.
Since, by hypothesis, $\psi $ belongs also to \ $L_{-\gamma }^{2}$ a theorem
by Agmon (see \cite{Agmon} Lemma 5.1) \ implies that $\psi \in H^{2,-\gamma
} $.\medskip

\textbf{(ii) }$\Rightarrow $\textbf{(iii) or (v).} Trivial.\medskip

\textbf{(iii) }$\Rightarrow $\textbf{(i). }From the fact that $G_{0}$ is the
right inverse of $-\Delta $ it follows that $\psi _{3}$ is a distributional
solution of the stationary Schr\"{o}dinger equation (\ref{Sch equation}).
Since $\psi _{3}$ belongs to $H^{2,-\gamma }\subset L_{-\gamma }^{2}$ the
claim follows.

\textbf{(v) }$\Rightarrow $\textbf{(i).} As before, we know that $\psi
_{5}\equiv \psi $ is a distributional solution of (\ref{Sch equation}). The
identity $\psi =G_{0}V\psi $ and a result by Ikebe (\cite{Ikebe}, Lemma 3.2;
see also \cite{TDMB}, Lemma 3.3) implies that $\left| \psi (\mathrm{x}%
)\right| $ $\leq C\ |\mathrm{x}|^{-1}$ as $|\mathrm{x}|\rightarrow \infty $,
provided that $V\in \mathrm{(I)}_{3}$. Taking into account the continuity of 
$\psi $, it follows that $\psi \in $ $L_{-\gamma }^{2}(\Bbb{R}^{3})$ for
every $\gamma >\frac{1}{2}$.%
\endproof%
\bigskip

\noindent \textbf{Remark. }We used the hypothesis $V\in \mathrm{(I)}_{3}$
only to prove that (v)$\Rightarrow $(i)$.$ All the remaining results hold
true under the weaker hypothesis that $V\in \mathrm{(I)}_{2}$.\bigskip

\noindent \textbf{Remark. }From the proof of the previous proposition it is
possible to extract the following \ result concerning the asymptotic
behavior and the local regularity of \ distributional solutions of the
stationary Schr\"{o}dinger equation: if $V\in \mathrm{(I)}_{2}$ and \ $\psi
\in $ $L_{-\gamma }^{2}(\Bbb{R}^{3})$ \ (with $\gamma >\frac{1}{2}$) \
solves equation (\ref{Sch equation}) then $\psi _{2}\in H^{2,-\gamma }\cap
C_{\infty }$.\bigskip

The ``equivalence theorem'' (Prop. \ref{Prop Equivalence}) shows that -- as
expected from the physical point of view -- the concept of (zero-energy)
resonance is largely independent from the technical tools needed to define
it. In this spirit, as pointed out by Enss \cite{En}, it would be
interesting to give a characterization of resonances which involves only the
time evolution and the position operator -- analogous to the RAGE
characterization of bound states and scattering states.

\end{document}